\documentclass[]{aastex}

\usepackage{emulateapj5}
\usepackage{onecolfloat}
\usepackage{graphicx}
\usepackage{fancyheadings}
\usepackage{ulem}
\usepackage{rotating}
\usepackage{lscape}

\def\arcsec{\hbox{$^{\prime\prime}$}}
\def\arcdeg{\mbox{$^\circ$}}%

\newcommand{\apl}{\lesssim}

\newcommand{\etal}{et al.}
\newcommand{\hI}{\mbox{${\rm H\ I}$}}
\newcommand{\kms}{\mbox{km\ s${^{-1}}$}}
\newcommand{\lya}{\mbox{${\rm Ly}\alpha$}}
\newcommand{\lyb}{\mbox{${\rm Ly}\beta$}}
\newcommand{\lyc}{\mbox{${\rm Ly}\gamma$}}

\def\N#1{{N({\rm #1})}}
\def\cm#1{\, {\rm cm^{#1}}}
\def\rtp{\, \right  ) }
\def\ltp{\left ( \,}
\def\perd{\;\;\; .}
 
\newcommand{\civof}{\ion{C}{4}$_{15}$}

\begin{document}

\twocolumn[%
\lefthead{Chen \etal}
\righthead{}

\slugcomment{Accepted for Publication in the Astrophysical Journal}

\title{ON THE ABSENCE OF WIND SIGNATURES IN GRB AFTERGLOW SPECTRA: CONSTRAINTS ON THE WOLF-RAYET WINDS OF GRB PROGENITORS}
\author{
HSIAO-WEN CHEN\altaffilmark{1}, 
JASON X.\ PROCHASKA\altaffilmark{2}, 
ENRICO RAMIREZ-RUIZ\altaffilmark{2,3},
JOSHUA S.\ BLOOM\altaffilmark{4,5}, 
MIROSLAVA DESSAUGES-ZAVADSKY\altaffilmark{6}, AND
RYAN J.\ FOLEY\altaffilmark{4}
}

%\newpage

\begin{abstract}

 We investigate available constraints on the circumstellar medium
(CSM) around long-duration $\gamma$-ray burst (GRB) progenitors from
afterglow spectra.  We first establish a statistical sample of five
GRB afterglow spectra that have been collected and analyzed with no
prior knowledge of the line-of-sight properties.  This sample is then
adopted for a uniform search of Wolf-Rayet wind signatures, as
represented by C\,IV $\lambda\lambda$ 1548, 1550 absorption doublets
at $\Delta\,v=-1000$ to $-5000$ \kms\ from the GRBs (hereafter
C\,IV$_{15}$).  We report the detection of a single C\,IV$_{15}$
absorber at $\Delta\,v\approx -1500$ \kms\ from GRB\,050730 and none
in the rest.  Our search yields an estimate of 20\% for the incidence
of C\,IV$_{15}$ absorbers with rest-frame absorption equivalent width
${\rm EW(C\,IV\,1548)}>0.2$ \AA\ near GRB host galaxies, consistent
with the incidence of intergalactic C\,IV$_{15}$ near classical damped
\lya\ absorbers toward quasar sightlines.  Including the two
C\,IV$_{15}$ absorbers previously known toward GRB\,021004, we further
demonstrate that the presence of H$^0$, C$^+$, and Si$^+$ together
with the absence of excited C$^+$ or Si$^+$ argue against a CSM
origin.  The null result is consistent with the expectation that the
circumburst medium is fully ionized by the afterglow radiation field.
We examine possible scenarios for the survival of the C$^{3+}$ ions,
including a clumpy wind model.  We find that a clumpy wind is unable to
effectively shield the ionizing radiation and allow C$^{3+}$ to
survive at $r<10$ pc from the afterglow.  We conclude that the lack of
CSM-originated C\,IV$_{15}$ absorbers is consistent with Wolf-Rayet
winds terminating at $< 30\ {\rm pc}$ from their progenitor stars.

\end{abstract}

\keywords{gamma rays: bursts---ISM: abundances---ISM:
kinematics---intergalactic medium}
]
\altaffiltext{1}{Department of Astronomy \& Astrophysics, University of
Chicago, Chicago, IL 60637, {\tt hchen@oddjob.uchicago.edu}}

\altaffiltext{2}{UCO/Lick Observatory; University of California, Santa
  Cruz, Santa Cruz, CA 95064, {\tt xavier@ucolick.org}}

\altaffiltext{3}{Institute for Advanced Study, Olden Lane, Princeton, NJ 08540}

\altaffiltext{4}{Department of Astronomy, 601 Campbell Hall, University of 
California, Berkeley, CA 94720 {\tt jbloom@astron.berkeley.edu}}

\altaffiltext{5}{Sloan Research Fellow}

\altaffiltext{6}{Observatoire de Gen\`eve, 51 Ch. des Maillettes,
        1290 Sauverny, Switzerland}
%\newpage

\section{INTRODUCTION}

\setcounter{footnote}{0}

A massive star origin of most long-duration, soft-spectrum
$\gamma$-ray bursts (GRBs) has now been well-established on empirical
grounds (see Woosley \& Bloom 2006 for a recent review).  Not only the
locations of these GRBs trace closely the spatial distribution of
young stars in their host galaxies (e.g.\ Bloom, Kulkarni, \&
Djorgovski 2002, Hammer \etal\ 2006; Fruchter et al.\ 2006), but a
growing number of afterglows are also observed to exhibit spectral
features that are characteristic of a core-collapse supernova (Hjorth
\etal\ 2003; Stanek \etal\ 2003; but see Fynbo \etal\ 2006a, Gal-Yam
\etal\ 2006, \& Della Valle \etal\ 2006).  In the popular collapsar
model, likely candidate progenitors of long-duration GRBs are
Wolf-Rayet stars that undergo a significant mass loss through stellar
winds (MacFadyen \& Woosley 1999).  It is therefore expected that the
circumstellar medium (CSM) has been regulated by the ejected wind
material throughout the lifetime of the progenitor star (Wijers 2001;
Ramirez-Ruiz \etal\ 2001; Chevalier \etal\ 2004; Ramirez-Ruiz \etal\
2005; van Marle \etal\ 2005; Eldridge \etal\ 2006; van Marle \etal\
2006).

In principle, the presence of Wolf-Rayet winds in the circumburst
medium can be demonstrated in an early-time, single-epoch afterglow
spectrum.  Specifically, ultraviolet spectra of local Wolf-Rayet stars
exhibit prominent P-cygni profiles in N\,V $\lambda\,1242$, Si\,IV
$\lambda\,1442$, and C\,IV $\lambda\,1550$ trannsitions (e.g.\
Niedzielski \& Rochowicz 1994; Willis \etal\ 2004) that indicate a
wind terminal velocity up to $v_{\rm term}=5000$ \kms\ (Figure 1).
Numerical simulations of single stars show that a free-expanding wind
will imprint a narrow absorption-line component in the afterglow
spectrum, blue-shifted ($\Delta\,v \apl -1000$ \kms) from the host
redshift (van Marle, Langer, \& Garc\'ia-Segura 2005).  Such
blue-shifted, high-velocity absorption components have been reported
in C\,IV transitions for GRB\,021004 at $z_{\rm GRB}=2.329$ (M{\o}ller
\etal\ 2002; Mirabal \etal\ 2003; Schaefer \etal\ 2003; Fiore \etal\
2005; Fynbo \etal\ 2005; Starling \etal\ 2005; see also Lazzati \etal\
2006), GRB\,030226 (Klose \etal\ 2004), and GRB\,050505 at $z_{\rm
GRB}=4.2741$ (Berger \etal\ 2006), but the origin of these
high-velocity C\,IV absorption components is ambiguous for three main
reasons.

First, over a large velocity interval, between $\Delta\,v=1000$ and
$\Delta\,v=5000$ \kms\ {\it blueward} of the host redshift, the
likelihood of finding an intervening absorber originated in a
foreground galaxy along the line of sight is non-negligible.  Second,
the presence of an intense ultraviolet radiation field from the
optical afterglow substantially increases the ionization of the
circumburst medium.  It therefore prohibits most ions from surviving
at $\le\,30$ pc radius from the afterglow. A CSM origin of these
absorbers would require the Wolf-Rayet winds to reach beyond 30 pc,
but observations of local Wolf-Rayet stars show that the wind blown
bubbles have a typical size of $<\,10$ pc (Gruendl \etal\ 2000).
Finally, the circumburst medium is expected to be highly ionized even
at $>\,30$ pc, due to the presence of the intense afterglow radiation
field.  But the observed C\,IV abundance alone is insufficient for
constraining the ionization state of the gas around the afterglow.  A
full consideration of all species at different ionization stages is
necessary (e.g.\ Fynbo et al.\ 2005).

We have established a statistical sample of five GRB afterglows for
which the afterglow spectra were collected and analyzed with no prior
knowledge of the absorption-line properties along the line of sight.
We have also assembled an early sample of previous GRB afterglows for
which the spectra are available to us.  The statistical sample allows
an unbiased estimate of the incidence of the blue-shifted,
high-velocity C\,IV absorption components.  The statistical sample and
early sample together allow us to investigate the nature of the
blue-shifted, high-velocity C\,IV absorption components based on a
suite of ionization models.  In \S\ 2, we present the GRB samples and
summarize the available spectroscopic data for each burst.  In \S\ 3,
we examine the possible intergalactic nature of these C\,IV absorbers
based on the relative abundances between different ionization states
of carbon and silicon.  In \S\ 4, we consider the afterglow radiation
field estimated from light-curve observations, and constrain the
extent of Wolf-Rayet winds based on the presence and absence of
high-velocity C\,IV absorbers in the afterglow spectra.  We further
discuss the scenarios that will allow C$^{3+}$ to survive the intense
afterglow radiation field.  Finally, we summarize our study in \S\ 5.

\section{THE GRB SAMPLES}

We have compiled a {\it statistical sample} of GRB sightlines for
which we resolve the C\,IV $\lambda\lambda\,1548, 1550$ absorption
doublet (i.e.\ a spectral resolution $\delta\,v < 180$ \kms) at
signal-to-noise ratios of $S/N > 7$ per resolution element so that the
C\,IV $\lambda\,1548$ transition would be observed at $> 3\,\sigma$
confidence level to an EW limit of 0.4 \AA\ over a 170 \kms\ velocity
interval.  The GRB sightlines in this sample are selected entirely
based on the data quality (spectral resolution $\delta\,v$ and $S/N$)
from the GRAASP\footnote{http://www.graasp.org/} database with no
prior knowledge of the presence or absence of intervening absorbers.
These include GRBs 050730, 050820, 050908, 051111, and 060418, and
form a statistically unbiased sample for a blind search of intervening
C\,IV absorbers.

In addition, we examine the GRB spectra that are available in the
literature, in search of Wolf-Rayet signatures.  These sightlines,
despite incomplete and biased by known light-of-sight properties,
offer additional insights for constraining the Wolf-Rayet nature of
GRB progenitor stars.  The GRBs in this catagory are GRBs 000926,
021004, and 030323.  We call this an ``early sample'', because the
events occurred prior to 2004.

In both the {\it statistical} and the {\it early} samples, we search
for blue-shifted, high-velocity components of C\,IV within
$|\Delta\,v|\le v_{\rm term}$ \kms\ in the afterglow spectra.  The
velocity threshold $v_{\rm term}$ is chosen to match the terminal
velocity observed for galactic Wolf-Rayet stars (Figure 1).  Based on
131 objects in the VIIth catalogue of galactic Wolf-Rayet stars by van
der Hucht (2001), we show that all have a terminal wind velocity of
$\le 5000\,\kms$.  The search of high-velocity C\,IV components is
therefore carried out at $|\Delta\,v|< v_{\rm term}=5000$ \kms.  An
additional selection threshold for $\Delta\,v = -1000$ \kms\ is
applied based on the maximum velocity of galactic winds observed in
starburst galaxies at $z>1$ (e.g.\ Adelberger \etal\ 2005).  The
nature of C\,IV within $\Delta\,v>-1000$ \kms\ becomes more ambiguous
due to possibility of a large-scale galactic wind.  We therefore
exclude this region from consideration.

The blue-shifted, high-velocity C\,IV absorbers identified over the
velocity interval of $-5000 \le \Delta\,v\le -1000$ \kms\ (hereafter
designated as C\,IV$_{15}$) from the GRB host redshift $z_{\rm GRB}$
represent likely wind signatures of the progenitor stars.  We also
search for the presence of C\,II, Si\,II, Si\,IV, and additional ions
whenever the transitions are covered.  These additional transitions
are crucial for constraining the ionizing radiation field around the
absorbing gas.  Here we summarize available spectroscopic data and our
search results.  For each source, the zero relatively velocity
$\Delta\,v=0$ corresponds to the redshift of the afterglow, which is
determined based on the absorption profiles of low-ionization
transitions such as C\,II, Si\,II, Al\,II, and Fe\,II.

\subsection{The Statistical Sample}

\subsubsection{GRB\,050730 at $z=3.96855$}

This GRB was first discovered by the Swift satellite on UT 2005 July
30 (Holland \etal\ 2005).  An optical source was found promptly using
the Ultraviolet-Optical Telescope (UVOT) on board Swift with $V =
17.6$ about three minutes after the burst trigger (Holland \etal\
2005).  This source was subsequently noted by Cobb \& Bailyn (2005) to
fade, confirming it as the optical afterglow.  We obtained an echelle
spectrum of the afterglow, using the MIKE spectrograph (Bernstein
\etal\ 2003) on the Magellan Clay telescope, four hours after the
initial trigger.  Descriptions of the data were presented in Chen
\etal\ (2005) and Prochaska \etal\ (2006).  The spectrum covers a full
spectral range from 3300 \AA\ through 9400 \AA\ with a spectral
resolution of $\delta\,v \approx 10$ \kms\ at wavelength
$\lambda=4500$ \AA\ and $\delta\,v \approx 12$ \kms\ at $\lambda=8000$
\AA.  The host of the GRB exhibits a strong damped \lya\ absorption
feature with $\log\,N(\hI)=22.15\pm 0.05$, and abundant heavy ions
with $>95$\% of the absorption confined within $-200\le\,\Delta\,v\le
50$ \kms.

Figure 2 shows the absorption-line profiles of C\,IV
$\lambda\lambda\,1548, 1550$, Si\,IV $\lambda\lambda\,1393, 1402$,
C\,II $\lambda\,1334$, and Si\,II $\lambda\,1526$ observed in the host
ISM.  The spectral region near the C\,IV absorption doublet is
contaminated by the atmosphere A-band absorption forest, but we
identify a C\,IV$_{15}$ absorber at $\Delta\,v=-1500$ \kms.
Comparison with a quasar spectrum obtained using the same instrument
at the same site (shown as the magenta curve in Figure 2) confirms the
lack of atmosphere absorption transition at the location of these
C\,IV$_{15}$ lines.  The identification is further supported by a
consistent kinematic profile of the Si\,IV $\lambda\lambda\,1393,
1402$ doublet transitions, for which we measure $\log\,N({\rm
Si\,IV})=13.9\pm 0.1$.  For the C\,IV transitions, we measure a lower
limit to the rest-frame absorption equivalent width of
EW$(\lambda\,1550)\ge 0.36$ \AA.  At this redshift, the C\,II
$\lambda\,1334$ transition is contaminated by a pair of C\,IV
absorption doublet at lower redshifts, $z=3.254$ and $z=3.258$.
Therefore, no constraints can be obtained.  We do, however, identify
weak \ion{Si}{2} $\lambda\,1526$ absorption and a strong \ion{Si}{2}
$\lambda\,1260$ profile.

\subsubsection{GRB\,050820A at $z=2.6147$}

This GRB was first discovered by the Swift satellite on UT 2005 August
20 (Page \etal\ 2005).  An optical transient, reported less than 1
hour after the GRB, was identified in data taken shortly after the
trigger (Fox \& Cenko 2005; Vestrand \etal\ 2006).  We obtained an
echelle spectrum of the afterglow, using the HIRES (Vogt \etal\ 1994)
on the Keck I telescope, an hour after the initial trigger.
Descriptions of the data were presented in Prochaska \etal\ (2006).
The spectrum covers a full spectral range from 3800 \AA\ through 8000
\AA\ with a spectral resolution of $\delta\,v \approx 7.5$ \kms\
across the entire spectral range.  The host of the GRB exhibits a
strong damped \lya\ absorption feature with $\log\,N(\hI)=21.0\pm
0.1$, and abundant heavy ions with $>95$\% of the absorption confined
within $-250\le\,\Delta\,v\le 200$ \kms.

Figure 3 shows the absorption-line profiles of C\,IV
$\lambda\lambda\,1548, 1550$, Si\,IV $\lambda\lambda\,1393, 1402$,
C\,II $\lambda\,1334$, Si\,II $\lambda\,1260$, and Al\,II
$\lambda\,1670$ observed in the host ISM.  While the host ISM shows
complex, multiple-component absorption features at $|\Delta\,v|<250$
\kms\ in C\,IV and Si\,IV transitions, it does not exhibit additional
blue-shifted, high-velocity component beyond this velocity range at a
3-$\sigma$ upper limit of EW($\lambda\,1548$)$=0.2$ \AA\ over
$\delta\,v=7.5$ \kms.

\subsubsection{GRB\,050908 at $z=3.344$}

This GRB was discovered by the Swift satellite on UT 2005 September 8
(Goad \etal\ 2005).  An optical transient, seen in data taken 14
minutes after the burst, was reported $\approx\,80$ minutes after the
GRB trigger (Torii et al. 2005; see also Cenko \etal\ 2005 and Li
\etal\ 2005).  We obtained a low-resolution spectrum of the afterglow,
using GMOS on the Gemini north telescope, four hours after the initial
trigger (Foley \etal\ 2005).  A moderate-resolution spectrum was taken
four hours later by our group using DEIMOS (Faber \etal\ 2003) on the
Keck II telescope (Prochaska \etal\ 2005).  The final stacked GMOS
spectrum covers a spectral range from 5073 \AA\ through 7945 \AA\ with
chip gaps at 6006--6022 \AA\ and 6972--6987 \AA\ a spectral resolution
of $\delta\,v\approx 200$ \kms.  The final stacked DEIMOS spectrum
covers a spectral range from 6335 \AA\ through 8960 \AA\ with a
spectral resolution of $\delta\,v \approx 40$ \kms.

The GMOS spectrum covers the redshifted hydrogen \lya\ $\lambda\,1215$
transition.  We measure $\log\,N(\hI)=19.2\pm 0.2$.  Figure 4 shows
the absorption-line profiles of C\,IV $\lambda\lambda\,1548, 1550$,
Si\,IV $\lambda\lambda\,1393, 1402$, C\,II $\lambda\,1334$, Si\,II
$\lambda\,1526$, and Al\,II $\lambda\,1670$ observed in the host ISM.
The host galaxy also exhibits abundant heavy ions with $>95$\% of the
absorption confined within $-300\le\,\Delta\,v\le 200$ \kms.  We do
not detect C\,IV$_{15}$ beyond this velocity range at a 3-$\sigma$
upper limit of EW($\lambda\,1548$)$=0.28$ \AA\ over $\delta\,v=40$ \kms.

\subsubsection{GRB\,051111 at $z=1.54948$}

This GRB was discovered by the Swift satellite on UT 2005 November 11
(Sakamoto \etal\ 2005).  An optical transient was found 27 s after the
burst on the ground (Rykoff \etal\ 2005).  An echelle spectrum was
taken an hour after the initial burst by the Keck Observatory staff,
using HIRES on the Keck I telescope.  However, the echelle spectrum
does not extend below 4000 \AA, where the redshifted C\,IV absorption
is expected.  We also obtained a low-resolution spectrum of the
afterglow, using GMOS on the Gemini north telescope, 2.6 hours after
the initial trigger.  The final stacked GMOS spectrum covers a
spectral range from 3800 \AA\ through 4400 \AA\ with a spectral
resolution of $\delta\,v \approx 170$ \kms.  Figure 5 shows the
absorption profiles of C\,IV $\lambda\lambda\,1548, 1550$, Si\,II
$\lambda\,1526$, and Al\,II $\lambda\,1670$ observed in the host ISM.
The host galaxy also exhibits abundant heavy ions with $>95$\% of the
absorption confined within $-200\le\,\Delta\,v\le 100$ \kms.  We do
not detect C\,IV$_{15}$ beyond this velocity range at a 3-$\sigma$
upper limit of EW($\lambda\,1548$)\,$=0.37$ \AA\ over $\delta\,v=170$
\kms.

\subsubsection{GRB\,060418 at $z=1.4901$}

This GRB was first discovered by the Swift satellite on UT 2006 April
18 (Falcone \etal\ 2006).  A candidate optical afterglow, with
$V=14.5$ mag, was reported from UVOT images taken starting 88 seconds
after the trigger (Falcone \etal\ 2006).  We obtained an echelle
spectrum of the afterglow, using the MIKE spectrograph on the Magellan
Clay telescope, 28 minutes after the initial trigger.  Descriptions of
the data were presented in Prochaska \etal\ (2006).  The spectrum
covers a full spectral range from 3500 \AA\ through 9000 \AA\ with a
spectral resolution of $\delta\,v \approx 12$ \kms\ at wavelength
$\lambda=7000$ \AA.  The host of the GRB exhibits abundant heavy ions
with $>95$\% of the absorption confined within $-200\le\,\Delta\,v\le
100$ \kms.

Figure 6 shows absorption-line profiles of C\,IV
$\lambda\lambda\,1548, 1550$, Si\,IV $\lambda\lambda\,1393, 1402$,
C\,II $\lambda\,1334$, Si\,II $\lambda\,1526$ and Al\,II
$\lambda\,1670$ observed in the host ISM.  Over the velocity interval
from $\Delta\,v=-5000$ \kms\ through $\Delta\,v=-1000$ \kms, we do not
detect additional components at a 3-$\sigma$ upper limit of
EW$(\lambda\,1548)=0.36$ \AA\ over a spectral resolution element of
$\delta\,v=12$ \kms.

\subsection{The Early Sample}

We have searched C\,IV$_{15}$ along the sightlines toward GRB\,000926
(Hurley \etal\ 2000; Harrison \etal\ 2001), GRB\,021004 (Shirasaki
\etal\ 2002; Holland \etal\ 2003), and GRB\,030323 (Graziani \etal\
2003; Gilmore \etal\ 2003), for which afterglow spectra are available
to us.  The afterglow spectrum of GRB\,000926 was obtained using the
Echelle Spectrograph and Imager (ESI; Sheinis \etal\ 2002) on the Keck
II telescope (Castro \etal\ 2003).  We found no C\,IV$_{15}$ at a
3-$\sigma$ upper limit to the rest-frame absorption equivalent width
EW$(\lambda\,1548)=0.35$ \AA\ over a spectral resolution element of
$\delta\,v=66$ \kms.  The afterglow spectrum of GRB\,030323 was
obtained using the FORS2 spectrograph on the VLT/UT4 telescope
(Vreeswijk \etal\ 2004).  We found no C\,IV$_{15}$ at a 3-$\sigma$
upper limit to EW$(\lambda\,1548)=0.04$ \AA\ over a spectral
resolution element of $\delta\,v=143$ \kms.

Afterglow spectra of GRB\,021004 were obtained, using various
spectrographs on multiple telescopes (Castander \etal\ 2002; Chornock
\& Filippenko 2002; Salamanca \etal\ 2002; Castro-Tirado \etal\ 2002;
M{\o}ller \etal\ 2002; Fiore \etal\ 2005; Starling \etal\ 2005).  The
host of this GRB does not exhibit a prominent damped \lya\ absorption
feature that is commonly seen in most GRB hosts (e.g.\ Vreeswijk
\etal\ 2004; Chen \etal\ 2005; Berger \etal\ 2006; Jakobsson \etal\
2006).  Based on the Lyman limit absorption discontinuity and the
observed absorption strength of the \lyb\ transition, Fynbo \etal\
(2005) estimated a neutral hydrogen column density of
$\log\,N(\hI)=19.5\pm 0.5$.  In addition, available echelle spectra of
the optical afterglow clearly resolve two C\,IV$_{15}$ absorption
components approaching $\Delta\,v=-3000$ \kms.  While a blue-shifted,
high-velocity component is expected in the circumstellar medium of a
Wolf-Rayet progenitor, the presence of two components is difficult to
explain under the free-expanding wind model of a single progenitor
star (e.g.\ van Marle \etal\ 2005).

We retrieved from the ESO Science Archive (program ID's 70.A-0599(B)
and 70.D-0523(D)) echelle spectra of the source obtained 13.5 hours
after the burst, using the Ultraviolet-Visual Echelle Spectrograph
(UVES; D'Odorico \etal\ 2000) on the VLT Kueyen telescope.  We
examined and reduced the individual spectra.  The data were weighted
by their S/N ratios and co-added to form a final stacked spectrum.
The final, stacked spectrum spanned a spectral range from 4200 to 9900
\AA\ with a spectral resolution of $\approx 3.8$ \kms\ per pixel, and
achieved a signal-to-noise ratio of S/N $\approx 20 $ per resolution
element at around 5200 \AA.  A copy of a low-resolution spectrum that
was obtained nearly 4 days after the GRB, using the Low Resolution
Imaging Spectrograph (LRIS) on the Keck I telescope was kindly made
available to us by N. Mirabal.  The spectrum covers the spectral range
around the redshifted \lya\ transition at the host redshift and has a
spectral resolution of $\approx 150$ \kms\ per pixel.  Descriptions of
the LRIS observations are presented in Mirabal \etal\ (2003).

We present in Figure 7 the absorption-line profiles of H\,I \lya\
$\lambda\,1215$, C\,IV $\lambda\lambda\,1548, 1550$, Si\,IV
$\lambda\lambda\,1393, 1402$, C\,II $\lambda\,1334$, Si\,II
$\lambda\,1526$, and Al\,II $\lambda\,1670$ observed in the host of
GRB\,021004.  Figure 7 shows that in addition to the main absorption
component at $|\Delta\,v|<400$ \kms\, two absorption components are
seen blue-shifted from the host redshift at $|\Delta\,v|=2700$
(Component 1) and 2900 \kms\ (Component 2), respectively.  A strong
\lya\ absorption feature is present at the location of the two
blue-shifted, high-veolicty C\,IV components, but not resolved in the
LRIS spectrum.  Mirabal \etal\ reported a rest-frame absorption
equivalent width of EW(\lya)$=3.9\pm 0.6$ \AA, which implies a total
neutral hydrogen column density\footnote{Mirabal et al. (2003) derived
progressively larger $N(\hI)$, from $\log\,N(\hI)=14.95$ to
$\log\,N(\hI)=16.11$, based on the observed EW(\lya), EW(\lyb), and
EW(\lyc).  These authors assumed that these transitions fall on the
linear part of the curve of growth, but noted the differences in the
derived $N(\hI)$ as suggestive of the lines being saturated.  The
differnce between our derived $N(\hI)$ and those from Mirabal \etal\
arises entirely due to the assumption these authors adopted, which is
invalid for saturated lines.} of $\log\,N(\hI) \sim 19.5$ for a
typical Doppler parameter of $b=20$ \kms.  Our $N(\hI)$ estimate is
consistent with that reported by Fynbo et al.\ (2005) from their HST
observations of the Lyman limit discontinuity.  We confirm this value
with a Voigt profile fit to the LRIS data and estimate an uncertainty
of 0.4\,dex which is dominated by systematics (continuum fitting and
line-blending).  In addition to the strong H\,I \lya\ transition, we
also observe abundant low-ions such as C$^+$ as evident by the C\,II
$\lambda\,1334$ transition at the location of Component 1.

We measure the ionic column density of each species using the apparent
optical depth method (Savage \& Sembach 1991).  The measurements are
presented in Table 1.  The large contrast in the observed relative
column densities of C$^+$ and C$^{3+}$ for components 1 and 2 suggests
that they arise in distinct phases (see \S\ 3.2).  Furthermore,
strong absorption from neutral hydrogen and low ionization species
such as C$^+$ strongly indicate that this absorber may not originate
in the highly ionized circumstellar medium of the progenitor.  We will
discuss various ionization scenarios in \S\ 3.2.

\section{THE NATURE OF BLUE-SHIFTED, HIGH-VELOCITY C\,IV ABSORBERS}

Our systematic search for C\,IV$_{15}$ absorbers along five GRB lines
of sight in the {\it statistical sample} has yielded identifications
of one absorber along the sightline toward GRB\,050730 at
$\Delta\,v\approx -1500$ \kms. Despite a heterogeneous spectral data
set, our survey is sensitive to C\,IV$_{15}$ absorbers of rest-frame
absorption equivalent width ${\rm EW(\lambda\,1548)}>0.04 -0.4$ \AA\
at the 3-$\sigma$ significant level, when scaled to a uniform
$\delta\,v=170$ \kms\ velocity resolution element.  We estimate that
the incidence of C\,IV$_{15}$ absorbers with ${\rm
EW(\lambda\,1548)}>0.4$ \AA\ is 20\% near a GRB afterglow.  

Despite a small sample, we attempt to determine the confidence level
of our estimate in the incidence of C\,IV$_{15}$ absorbers based on a
bootstrap re-sampling analysis.  We first adopt the five GRB
sightlines as the parent sample.  Then we randomly select five fields
from the parent sample, allowing for duplication.  Next, we measure
the incidence of C\,IV$_{15}$ absorbers in the new sample.  Finally,
we repeat the procedure 1000 times and measure the scatter in our
measurements.  Our bootstrap analysis yields a 68\% confidence
interval of $0-40$\% in the incidence of C\,IV$_{15}$ absorbers with
${\rm EW(\lambda\,1548)}>0.4$ \AA.

Including sources in the {\it early sample} from the literature (GRBs
000926, 021004, and 030323), we confirm additional two components
along the sightline toward GRB\,021004 at $\Delta\,v\approx -2800$
\kms.  We note that possible C\,IV$_{15}$ features have also been
reported for GRB\,030226 at $z_{\rm GRB}=1.986$ by Klose \etal\ (2004)
and discussed by Shih \etal\ (2006).  The high spectral resolution
data are not accessible to us.  The source is therefore not included
in our analysis.  A summary of the observed absorption line properties
is presented in Table 2, where we list for each field the estimated
isotropic energy release of $\gamma$-ray photons $E_{\rm iso}$, the
redshift of the GRB host, $N(\hI)$ and metallicity of the host ISM,
the maximum velocity span of the observed blue-shifted C\,IV
absorption doublet, and the time period when the afterglow spectra
were obtained $t_{\rm obs}^{\rm spec}$.  We find no correlation
between the presence of C\,IV$_{15}$ and other factors, such as
$E_{\rm iso}$ or $t_{\rm obs}^{\rm spec}$.

While a Wolf-Rayet wind from the GRB progenitor is a natural candidate
for explaining the observed C\,IV$_{15}$ absorbers, it is clear that
the majority of GRB afterglow spectra do not exhibit these features.
Here we investigate the nature of these C\,IV$_{15}$ clouds, taking
into account the presence of ions at lower ionization stages.

\subsection{The Likelihood of Contaminations by Intervening Absorbers at $z_{\rm abs}<z_{\rm GRB}$}

We first consider the statistical sample of five GRB lines of sight as
a whole and evaluate the likelihood of finding an intervening
C\,IV$_{15}$ absorber over $|\Delta\,v|=1000-5000$ \kms\ at $z_{\rm
CIV} < z_{\rm GRB}$ based on known statistics from quasar
absorption-line studies.  A non-negligible probability of finding a
foreground absorber over the redshift pathlength surveyed by this sample
would challenge the validity of attributing these absorbers to the GRB
host environment.

Random surveys toward quasar lines of sight have shown a mean number
density per unit redshift interval of $n_{\rm C\,IV}(z=2)=1.8\pm 0.4$
for C\,IV absorbers of ${\rm EW}(\lambda\,1548)>0.4$ \AA\ and $n_{\rm
C\,IV}(z=2)\sim 5$ for ${\rm EW}(\lambda\,1548)>0.15$ \AA\ 
(Steidel 1992).  The total redshift
pathlength over $\Delta\,v=1000-5000$ \kms\ along the GRB lines of
sight in our sample of five is $\Delta\,z=0.23$.  Therefore, the
probability of detecting at least one random C\,IV absorber of ${\rm
EW}(\lambda\,1548)>0.4$ is 34\%.  At lower threshold ${\rm
EW}(\lambda\,1548)>0.15$, the probability of detecting at least one random
absorber at lower redshifts is 64\%.  We note that the C\,IV$_{15}$
absorber toward GRB\,050730 has ${\rm EW}(\lambda\,1548) > {\rm
EW}(\lambda\,1550)\ge 0.36$ \AA.

It is also necessary, however, to take into account the possibility of
large-scale clustering between C\,IV absorbers through clustering of
the absorbing galaxies.  Galaxy clustering may be important in this
analysis, particularly because the search is carried out near a GRB
host galaxy.  Unfortunately, the cross-correlation function of intervening C\,IV
absorbers and GRB host galaxies is not well determined.  Existing
two-point correlation measurements are for C\,IV absorbers of a wide
range of observed column density (e.g.\ Petitjean \& Bergeron 1994).
These are likely underestimates of the clustering extent and amplitude
for strong C\,IV absorbers, because stronger absorbers are expected to
cluster more strongly with one another (Adelberger \etal\ 2005).

As a first estimate, we have searched a sample of 53 quasar sightlines 
from the Keck/UCSD High Resolution database (Prochaska \etal\ 2007) 
with known intervening super Lyman limit systems (SLLS) of 
$\log\,N(\hI)>19$ or damped \lya\ absorbers (DLAs) of $\log\,N(\hI)\ge 
20.3$ (comparable to the $N(\hI)$ found in GRB host galaxies) at 
$1.7<z_{\rm abs}<4.3$.  The DLAs are believed to trace the majority of 
the galactic population at $z>2$.  Although their luminosity (or mass) 
function has not been established empirically, current expectation is 
that a significant fraction correspond to sub-$L_*$ galaxies (e.g.\ 
Weatherley \etal\ 2005) similar to GRB host galaxies (e.g.\ 
Christensen, Hjorth, \& Gorosabel 2004).

We searched for C\,IV$_{15}$ absorbers in the vicinity ($-5000\ {\rm km
s^{-1}} \leq \Delta v \leq -1000\ {\rm km s^{-1}}$) of intervening DLAs
toward background quasars (Figure 8).  This QSO-DLA sample serves as a
control sample for estimating the statistical significance of possible
``over-abundance'' of C\,IV$_{15}$ absorbers in the vicinity of GRB
host galaxies.  In the sample of 53 DLAs identified along quasar
sightlines, we find a total of six C\,IV$_{15}$ absorbers to an EW
limit of 0.2 \AA\ near five DLAs.  Therefore, the incidence of
C\,IV$_{15}$ from a DLA is $\approx 0.11$.  Namely, there is on
average one C\,IV$_{15}$ absorber in the vicinity of every nine DLAs.
Adopting this expectation, we estimate that the probability of
detecting at least one C\,IV$_{15}$ absorber near a GRB host galaxy in
our {\it statistical sample} of five GRBs is 42\%.  While the sample
of GRB host galaxies is still small, our exercise shows that the
probability for the C\,IV$_{15}$ absorber toward GRB\,050730 to
originate in a foreground galaxy at $z<z_{\rm GRB}$ is non-negligible.

\subsection{Constraints from the Absence of Excited Ions}

A nearly generic feature in the afterglow spectra of the ISM
surrounding the GRB is the presence of the fine-structure absorption
of C$^+$, Si$^+$, O$^0$ and Fe$^+$ (e.g.\ Chen \etal\ 2005; Berger
\etal\ 2006; Fynbo \etal\ 2006).  While these lines were initially
interpreted as signatures of a high density, circumstellar medium,
Prochaska, Chen, \& Bloom (2006; hereafter PCB06) demonstrated that
indirect UV pumping would dominate the excitation if the gas is at
$r\apl 100$\,pc of the afterglow.  Therefore, the presence or absence
of absorption due to excited ions constrains the distance of the gas
from the GRB event (see also Shin et al.\ 2006; Vreeswijk \etal\ 2006).

We examine the velocity profiles of various ions associated with the
C\,IV$_{15}$ absorbers from GRB~021004 and from GRB~050730 (Figure 9).
We include the low-ionization transitions of C\,II $\lambda\,1334$ and
Si\,II$\lambda\,1260$ for GRB\,021004 and GRB\,050730, respectively,
together with their associated fine-structure transitions.  Both C$^+$
and Si$^+$ have a $J=1/2$ ground-state and a corresponding $J=3/2$
excited fine-structure level.  It is clear that the ground-state is
signficantly populated for Component~1 of GRB~021004 and for the
majority of the line-profile for GRB~050730.  We set upper limits to
the ratio of the excited to ground level $f \equiv N_{J=3/2} /
N_{J=1/2}$ of $f< 0.05$ for Component~1 of GRB~021004 and $f < 0.1$
for GRB~050730.  Although optical depth effects will generally imply
$f<1$ (PCB06), the very low $f$ values require the gas lay at large
distance from the GRB afterglow.

A lower limit to the distance of the gas from the afterglow can be
obtained by calculating the total number of excitations from the
afterglow prior to the onset of the spectroscopic observations,
$t^{\rm spec}_{\rm obs}$.  For both of these GRBs, we can neglect
de-excitations from spontaneous decay because the lifetimes of the
excited levels are considerably longer than $t^{\rm spec}_{\rm obs}$.
Similarly, if the electron density of the gas were sufficient to
collisionally de-excite the states, one would predict $f \gg 0.1$
(Silva \& Viegas 2000).  Adopting GRB~021004 as an example, we find
from Figure 12 that the optical afterglow light curve is very
complicated for this GRB (e.g.\ Pandey \etal\ 2003; Bersier \etal\
2003).  We therefore derive a spline-fit model of the empirical
$R_C$-band measurements reported by Holland \etal\ (2003) and Uemura
\etal\ (2003).  In addition, we assume that $F_\nu \propto \nu^\beta$
with $\beta = -1$ noting that our calculation is not sensitive to this
choice.  We must also model the absorption line-profile of the gas
`cloud' to determine the number of photons absorbed prior to the onset
of our spectroscopic observations.  A single absorber with column
density $N({\rm C\,II}) = 10^{14.5} \cm{-2}$ and a Doppler parameter
$b = 10$ \kms\ is a conservative approximation to the observed
line-profile (the equivalent width actually underestimates the observed
value).  In the following, we only consider the
\ion{C}{2}~$\lambda\lambda$1036 and 1334 transitions.

The number of photons absorbed from ions in the ground-state of C$^+$
prior to spectroscopic observations is estimated by integrating over
the light curve from $t_i=1000$ s to $t^{\rm spec}_{\rm obs}$.  We
choose $t_i = 1000$ s because the light curve is not well defined
prior to this time and to give a more conservative estimate of the
photon number.  Integrating to $t_i = 0$ s would likely increase the
photon number by $<30\%$.  We find a photon surface density at
distance $r$ from the afterglow,
\begin{equation}
N_{\rm phot} = 10^{18}  \ltp \frac{{\rm 10\,pc}}{r} \rtp^2 \; {\rm cm^{-2}} \perd
\end{equation}
While only a fraction of the excitations would be followed by a
spontaneous decay to the $J=3/2$ level in the ground term, a
conservative estimate is $50$\%, because the upper level has a higher
$J$ value and correspondingly higher statistical weight.  To set a
lower limit to the distance of the gas from the afterglow, we
calculate the distance at which the photon surface density matches
twice the column density limit for the $J=3/2$ level of C$^+$ as
observed from the UVES observations, $N(\rm{C\,II^*}) < 10^{13}
\cm{-2}$.  This implies the gas occurs at a distance $r_{\rm min} > 1$
kpc from the GRB afterglow.  We derive a similar value for the Si$^+$
gas associated with GRB~050730.

The above argument is based primarily on the direct observations of
the afterglow flux.  No extrapolation is required for the prompt
emission or to higher energies than observed for the afterglow
radiation field.  Furthermore, the conclusions are based on empirical
observations of the abundances and are largely independent of
shielding scenarios (which we will consider to explain the survival of
C$^{3+}$; \S\ 4.3).  This is understood by the fact that
optical depth effects would yield excitation of C$^+$ in a thin layer
of the gas facing the afterglow, producing a non-negligible amount of
C\,II$^*$ transitions.  For example, in a layer of C$^+$ ions with
$\N{C^+} = 10^{13} \cm{-2}$ where the \ion{C}{2} transitions are
optically thin, excited C$^+$ ions are expected to dominate but are
absent in our data.

However, it is possible to design a scenario in which layers of
high-column density gas absorb a large fraction of $N_{\rm phot}$ and are
consequently photo-ionized by the afterglow just prior to $t^{\rm
spec}_{\rm obs}$.  This would require significant fine-tuning.
Altogether, the absence of fine-structure absorption lends strong
support for the conclusion that the C and Si ions identified at $v
\approx 2800$ \kms\ and $v \approx 1500$ \kms\ toward GRB\,021004 and
GRB\,050730, respectively, are not circumstellar material of the GRB
progenitor.

\subsection{Are the Absorbers Photo-ionized by the Metagalactic Radiation Field?}

If the C\,IV$_{15}$ absorbers found along the sightline toward
GRB\,021004 originate in foreground clouds at $z_{\rm C\,IV} < z_{\rm
GRB}$, then these gaseous clouds are expected to be photo-ionized
predominantly by the ultraviolet background radiation or possibly a
local starburst (Simcoe 2006).  Observations of abundance ratios
between different ionization stages of a given species can be applied
to constrain the ionization state of the gas.

We use the Cloudy software (Ferland \etal\ 1998; version 06.02) to
calculate the expected population ratios between different ionization
stages of carbon and silicon for clouds of plane parallel geometry and
under photo-ionization equilibrium.  We have assumed a metallicity of
0.1 solar and $\log\,N(\hI)=15$, appropriate for intervening, strong
C\,IV absorbers along quasar lines of sight (e.g.\ Cowie \etal\ 1995).
The results are not sensitive to the adopted metallicity and $N(\hI)$.
We also note that the assumed solar relative abundances are not
relevant because we compare pairs of ions for the same
elements\footnote{Our analysis differs from those presented in Fiore
et al. (2005) in two important aspects.  First, we adopt only lower
limits for saturated transitions.  Second, we compare the same
elements, but in different ionization stages.  This is necessary due
to potential biases imposed by different chemical abundance patterns
in the CSM.}.  Figure 10 shows the abundance ratios of C$^+$ to
C$^{3+}$ and Si$^+$ to Si$^{3+}$ versus the ionization parameter
$U\equiv \phi_{\gamma}/c\,n_{\rm H}$, which represents the number of
ionizing photons available per hydrogen particle.

Column density measurements presented in Table 1 for both
high-velocity C\,IV components allow us to place limits on the
ionization parameter.  For component 1, C\,II $\lambda\,1334$ is
saturated.  We can therefore place an upper limit at $\log\,U=-2.9$.
For component 2, both C\,IV and Si\,IV are saturated.  The
(C\,II/C\,IV) and (Si\,II/Si\,IV) ratios yield a consistent lower
limit of $\log\,U>-2.1$.  We note that $\log\,U \apl -3$ is relatively
low, even for Lyman limit systems (e.g.\ Bergeron \& Stas\'inska 1986;
Lopez \etal\ 1999; Chen \& Prochaska 2000).  On the other hand, the
C\,II/C\,IV ratio observed for the component 2 is typical of what is
obseved in the intergalactic medium (e.g.\ Songaila 2006).

The inferred limits for the ionization parameter $U$ allow us to
examine the gas properties of the absorbers.  Adopting a meta-galactic
radiation field of $J_{912}=1.4\times 10^{-21}$ erg s$^{-1}$ cm$^{-2}$
Hz$^{-1}$ sr$^{-1}$ measured at $z\sim 2$ (Scott \etal\ 2000), we
derive a gas density of $n_{\rm H}\sim 0.01-0.1$ cm$^{-3}$ for
$\log\,U=-2$ to $\log\,U=-3$.  Comparisons between observations and
expectations of photo-ionized gas in the intergalactic medium
therefore demonstrate that the observed relative ionic abundances for
the two C\,IV components can be naturally explained under a
photo-ionization equilibrium with the meta-galactic radiation field.
In particular, the upper limit on the ionization parameter for
component 1 indicates that a higher gas density and lower ionization
intensity is favored.  Our photo-ionization analysis substantiates the
fine-structure analysis from the previous section in that the gas is
located at large distance from the GRB afterglow.

If we are to interpret the \civof\ absorption associated with
GRB~021004 as intergalactic, then one might expect to identify a
galaxy associated with this gas.  Fynbo \etal\ (2005) find six faint
sources in HST imaging in a 10\arcsec$\times$10\arcsec\ region around
the host.  This overdensity of faint objects is qualitatively
consistent with the overabundance of Mg\,II absorbers at lower
redshift along the sightline.  Fiore et al. (2005) reported two groups
of Mg\,II absorbers identified at $z=1.380$ and $z=1.602$ along the
sightline (see also Prochter \etal\ 2006).  Figure 11 shows these
optical images obtained using HST/ACS with the F606W filter three days
(left panel) and eight months (right panel) after the trigger.  The
images are retrieved from the HST data archive (Program ID $=$ 9405).
While the afterglow was still bright in the first epoch image, we
point out a faint companion at an angular distance
$\Delta\,\theta=0.3$\arcsec\ to the host galaxy, which corresponds to
2.5 kpc projected distance at $z=2.299$.  The host galaxy and the
companion together within a 0.5\arcsec-radius aperture have a total
brightness of $AB({\rm F606W})=24.55\pm 0.05$ (Fynbo \etal\ 2005).  We
estimate that the companion contributes roughly $1/10$ of the total
flux and therefore has $AB({\rm F606W})\approx 27.3$.

Recall that a \lya\ absorption line with ${\rm EW}=3.9\pm 0.6$ \AA\ is
also identified at the location of the two C\,IV$_{15}$ components
(\S\ 2.2), although unresolved.  The presence of \lya\ absorber of
$\log\,N(\hI)=19.5$ presents a further challenge to the origin of the
absorber arising in the host environment.  On the other hand, the
\lya\ absorption strength is consistent with originating in the
extended gaseous halo of a faint dwarf galaxy at lower redshift (Chen
\etal\ 2001).  No deep image of the field around GRB\,050730 has been
reported yet.

\section{IMPLICATIONS FOR THE CIRCUMSTELLAR MEDIUM OF GRB PROGENITOR STARS}

The survey results from a statistical sample show that the majority of
afterglow spectra do not exhibit expected wind signatures from
Wolf-Rayet progenitors, consistent with the expectation that the
circumburst medium is highly ionized.  In addition, the fine-structure
and photo-ionization analysis show that the known C\,IV$_{15}$
absorbers near GRB afterglows most likely originate in a foreground
galaxy along the sightline.  Here we take into account the known
radiation field from afterglow light-curve observations.  The presence
of wind signatures in an afterglow spectrum determines the minimum
distance of the absorbing ions to the afterglow.  The absence of wind
signatures determines the maximum distance reached by the wind prior
to the GRB phase, if the absence is due to photo-ionization of the
wind.  Considering a sample of afterglow spectra together allows us to
estimate the typical extent of Wolf-Rayet winds.  In the following
discussion, we consider both smooth and clumpy media for allowing
effective shielding of ionizing photons from the afterglow.

\subsection{The Extent of Wolf-Rayet Wind around the Progenitor Stars}

Figure 12 shows the optical light curves of four GRB afterglows in our
spectroscopic sample, for which early-time ( from $<$ 10 minutes after
the initial burst) photometric measurements are available.  All but
one of these afterglows reached an initial brightness of $R <15.5$ mag
and declined by 2 magnitudes in less than two hours (with the
exception of GRB\,021004).  Even at $R=17.5$, the corresponding
intrinsic luminosity easily competes with the brightest quasars known
at this redshift (see e.g.\ Fan \etal\ 2001).  Despite a brief
lifetime, the afterglow can easily ionize the circumstellar medium to
large radii (see PCB06).  The presence or absence of the C\,IV$_{15}$
doublet features, coupled with the radiation field measured from
afterglow observations, therefore, determines the extent of the
Wolf-Rayet wind from the progenitor stars.

To assess the impact of the afterglow over the circumstellar medium,
we estimate the number of ionizing photons released to the surrounding
medium prior to the onset of the spectroscopic observations.  We first
adopt the afterglow of GRB\,050820 as an example, given its well
sampled optical light curve that shows both the prompt emission at
$t_{\rm obs} < t_{\rm GRB} + 5$ min and afterglow radiation at later
times (Vestrand \etal\ 2006).
%We estimate an isotropic release
%of $\phi_{\gamma}(h\nu=13.6 - 27.2\,{\rm eV})\approx 7.2\times
%10^{60}$ photons and $\phi_{\gamma}(h\nu=8-13.6\,{\rm eV})\approx
%10^{61}$ photons before the HIRES observations took place
%$\Delta\,t=t_{\rm obs}^{\rm spec}=3240$ s.  
The transient nature of the afterglow indicates that accurate
constraints on the ionization fraction of the surrounding medium
require a time-dependent photo-ionization calculation (e.g.\ Perna \&
Lazzati 2002).

For an approximation of the ionization fraction of C$^{3+}$, however,
we note that the recombination rate coefficient $\alpha_R({\rm
C}^{4+})\sim 10^{-11.5}$ cm$^3$ sec$^{-1}$ (Nahar \& Pradhan 1997) at
a gas temperature of $T\sim 10^{4-5}$ K and electron density $n_e \sim
10^4$ cm$^{-3}$, typical of a compact H\,II region, yields a
characteristic recombination time scale
\begin{equation}
t_{\rm rec}({\rm C}^{4+})\sim\,1\,{\rm year} \gg\ t_{\rm obs}^{\rm spec}.
\end{equation}
Since the timescale for observation ($<$ a few hours in the frame of
the medium external to the GRB outflow) is much smaller than the
recombination time for each GRB event (Table 2), the ionization
fraction of the circumstellar medium can be approximated by
considering only the amount of ionizing photons available to ionize
C$^{3+}$.

For the afterglow of GRB\,050820, we estimate an isotropic release of
total ionizing photon number, $\phi_{\gamma}(h\nu\ge 64\,{\rm
eV})\approx 1.8\times\,10^{60}$ over $\Delta\,t=t_{\rm obs}^{\rm
spec}=3240$ s.  Defining $r_{\rm min}$ as the radius of the ionization
front where 99.99\% of C$^{3+}$ are ionized, we have $n({\rm
C}^{3+})/n(C^{4+})\approx 10^{-4}$ in an optically thin regime or

\begin{equation}
\frac{n({\rm C}^{3+})}{n({\rm C}^{4+})}=10^{-4}=\exp\left[\frac{-\phi_\gamma\,\sigma_{ph}({\rm C}^{3+})}{4\pi r_{\rm min}^2}\right].  
\end{equation}
Therefore,
\begin{equation}
r_{\rm min}=\left[\frac{\phi_{\gamma}\,\sigma_{ph}({\rm C}^{3+})}{1.1\times\,10^{39}}\right]^{1/2}\,{\rm pc}.
\end{equation}
For $\sigma_{ph}({\rm C}^{3+})=6.6\times 10^{-19}$ cm$^2$, we find
that 99.99\% of C$^{3+}$ within 33\,pc would be ionized.  The lack of
C\,IV$_{15}$ in the spectrum is therefore consistent with the winds
terminating at $<\,30$ pc and becoming fully ionized shortly after the
GRB event.

The afterglow spectrum of GRB\,050820 presented in \S\ 2.1.2 is
sensitive to C\,IV transitions of $\log\,N_{2\,\sigma}({\rm C\,IV})\ge
13.7$ at $\ge\,2$-$\sigma$ significance level for a nominal Doppler
parameter $b=15$ \kms.  Therefore, it does not exclude the possibility
that Wolf-Rayet winds with $\log\,N({\rm C\,IV})< 13.7$ may be present
at $r>r_{\rm min}$.  The lack of wind signatures in the host of
GRB\,050820 implies that if such winds exist and contain a large
amount of C$^{3+}$ ($\log\,N({\rm C\,IV}) > 13.7$) then it did not
reach beyond 30\,pc.

%To constrain the extent of Wolf-Rayet wind of massive stars, we
%first revisit the light curves presented in Figure 12.  GRBs 021004
%and 050820 have the best-sampled light curves in our sample, and we
%adopt these two sightlines for comparison.  
The presence of the two C\,IV$_{15}$ components along the sightline
toward GRB\,021004 have been considered in the literature as the most
likely candidate for Wolf-Rayet winds in the circumburst environment
of GRBs, although our analyses in \S\ 3 show that at least one
(Component 1 at $\Delta\,v=-2675$ \kms) is best-explained by a
foreground cloud located at $> 1$ kpc from the afterglow and
photo-ionized by the ultraviolet background radiation field.

  Allowing the possibility for Component 2 at $\Delta\,v=-2900$ \kms\
to arise in the CSM of the progenitor star, we estimate the minimum
radius for the C$^{3+}$ ions to survive the afterglow radiation field.
Adopting the total number of ionizing photons estimated by Lazzati
\etal\ (2006) from both the prompt emission and the afterglow phase
$\phi_\gamma=4\times 10^{60}$, we derive a minimum radius $r_{\rm
min}\sim 50$ pc according to Equation (4) and find that more than
$99$\% of the C$^{3+}$ ions are ionized within this
radius\footnote{This is a factor of three smaller than the distance of
the C\,IV clouds derived by Lazzati et al.  The difference is
primarily due to a different ionization fraction adopted by these
authors.  Comparison between Equation (3) in the present paper and
Equation (11) in Lazzati et al. shows that these authors have allowed
the gas to be $\sim\,60$\% ionized when defining $r_{\rm min}$.  This
leads to $r_{\rm min}\sim 150$ pc.  We note that at this ionization
level the spectra cannot rule out the possibility that the observed
$N({\rm C\,IV})$ is due to gas at $r < 150$ pc, as it would only imply
a total gas column for carbon of $\log\,N({\rm C})\sim 15$ within this
radius.  It is therefore necessary to adopt a more stringent
ionization threshold.}.  Combining the upper limit derived for the
extent of Wolf-Rayet wind around the progenitor star of GRB\,050820
($< 30$ pc) with $r_{\rm min}$ derived for GRB\,021004, we find a
maximum extent of $30-50$ pc for the Wolf-Rayet winds around GRB
progenitor stars.

\subsection{Can the C$^{3+}$ Ions Survive the Afterglow Radiation Field at $r< 30$ pc?}
\label{sec:clump}

 The large distance implied for the Wolf-Rayet winds of the progenitor
of GRB\,021004 from observations of Component 2 at $\Delta\,v=-2900$
\kms\ is much larger than what is commonly observed for local
Wolf-Rayet stars.  The typical wind terminal shocks are found at $r
\apl 10$ pc (Garc\'ia-Segura \etal\ 1996; Gruendl \etal\ 2000).
%The strong C\,IV components found in the
%spectra of GRB\,021004 with $\log\,N({\rm C\,IV})>14$ suggests that if
%these absorbers arise in the circumstellar medium of the progenitor
%star, then the Wolf-Rayet wind would likely reach beyond 60 pc at
%$v_{\rm term}\approx 3000$ \kms, as the C$^{3+}$ ions would still be
%95\% ionized at $r<55$ pc.  On the other hand, observations and
%theoretical models of local Wolf-Rayet stars show a typical wind
%terminal shock radius of $\apl 10$ pc (Garc\'ia-Segura \etal\ 1996;
%Gruendl \etal\ 2000).  Therefore, the inferred radius of the wind
%terminal shock for GRB progenitor stars appears to be much larger than
%what is commonly observed around Wolf-Rayet stars.
In an effort to understand the physical conditions that determine the
location of the wind terminal shock around a GRB progenitor star, van
Marle \etal\ (2006) explored the parameter space spanned by the ISM
density of the GRB host, the wind velocity, and the mass-loss rate
during the Wolf-Rayet phase of the progenitor.  According to their
calculation, for the wind to reach beyond 50 pc at a wind-speed of
$v_{\rm WR}=3000$ \kms\ only a narrow parameter space is allowed.
Their Figure 12 (van Marle \etal\ 2006) shows that a mass-loss rate of
$1\times 10^{-5}\,{\rm M}_\odot\,{\rm yr}^{-1}$, an ISM density
$\log\,\rho_{\rm ISM} < -24\ {\rm g}\,{\rm cm}^{-3}$ or $\apl 1\,{\rm
cm}^{-3}$ need to be satisfied.  A lower ISM density or higher
mass-loss rate allows the wind to reach large distances.

A simple analytical treatment can give some useful insight. This is
because a strict upper limit on the radial extent of the Wolf-Rayet
wind can be obtained by computing the radius of the termination shock
for a free expanding wind. The radius at the inner edge of the wind
bubble can be found by balancing the wind ram pressure with the
post-shock cavity pressure. For a star that loses mass at a rate
$10^{-5}\,\dot{M}_{-5}\, M_\odot\, {\rm yr}^{-1}$ with a wind velocity
$10^{3}\,v_{w,3}$ km s$^{-1}$ in interstellar gas with density $10^{0}\,
n_{{\rm ISM},0}\, {\rm cm}^{-3}$, we have an inner termination shock
radius
\begin{equation}
r_{\rm t}=15\,\dot{M}_{-5}^{3/10}\,v_{w,3}^{1/10}\,n_{{\rm
ISM},0}^{-3/10}\,t_6^{2/5}\ \rm{pc}.
\label{bubble}
\end{equation}
where $10^{6}\,t_6$ is the lifetime of the star in Myr. For sufficiently
high $\dot{M}$ or low $n_{\rm ISM}$, the termination shock radius
given in equation (\ref{bubble}) can become larger than $r_{\rm
min}$.  This requires
\begin{equation}
\dot{M}_{-5}^{3/10}\,v_{w,3}^{1/10}\,n_{{\rm ISM},0}^{-3/10} \gtrsim 3
\left({r_{\rm min} \over 50\;{\rm pc}}\right),
\label{bubble}
\end{equation}
which shows that if the wind is specially massive ($\dot{M} \geq
10^{-4}\,M_\odot\,{\rm yr}^{-1}$) or the surrounding pressure is low
($n_{\rm ISM} \leq 10^{-1}\,{\rm cm}^{-3}$), $r_{\rm t}$ falls above the
radial range for which 99.99\% of the C$^{3+}$ ions are
ionized. Interestingly, this low gas density is consistent with the
low $N(\hI)$ observed in the host galaxy of GRB\,021004,
$\log\,N(\hI)=19.5\pm 0.5$.

Here we consider possible scenarios that would help to alleviate the
burden of forcing the wind out to beyond 30 pc.

We first consider the structured jet model discussed in Starling
\etal\ (2005).  These authors proposed that the ionization of
circumstellar medium is dominated by a narrow and fast-moving jet, but
the line of sight encompasses a wider and slower-moving jet through
the wind, where the C$^{3+}$ ions can survive at smaller distances to
the burst.  We consider this scenario unlikely for two main reasons.
First, following this structured jet model, we expect to observe
partial coverage of the absorption transitions along the line of
sight.  The C\,IV $\lambda\,1548$ absorption line profile of Component
2 is completely saturated over the central $\Delta\,\lambda_{\rm
obs}=0.8$ \AA\ (Figure 7).  We constrain the covering factor of
C$^{3+}$ ions along the line of sight at $f>97$\% at the 2-$\sigma$
significance level.  Second, light curve observations and modelling of
GRBs 021004 and 050820 show that prompt emission of the GRB event
contributes only a small fraction to the total ionizing photons
(Vestrand \etal\ 2006; Lazzati \etal\ 2006).  Therefore, the afterglow
serves both as the background light probe and as the principle source
of ionization of the circumstellar medium.

Next, we consider the possibility of a clumpy wind for a chemical
composition that is characteristic of local Wolf-Rayet winds.
Wolf-Rayet type stellar winds are found to have significant
inhomogeneous spatial distributions and a chemical composition
dominated by helium and carbon with little hydrogen (Crowther 2007).
Dense clumps, located at smaller distances to the afterglow along the
line of sight can in principle serve as a natural screen for shielding
the clumps that are further away from the progenitor stars.  The same
model is commonly applied to study the presence of dusty tori in the
vicinity of quasars (e.g.\ K{\"o}nigl \& Kartje 1994).

  To test whether a clumpy wind is a viable solution, we use the
compressible turbulent hot-star outflow description of Moffat et
al.\ (1994) as a working model, in which a simple scaling relation
between the size, $l_{\rm c}$, and density, $\rho_{\rm c}$ of the
clumps is assumed to prevail: $\rho_{\rm c} \propto l_{\rm
c}^{-1}$. What is more, Moffat et al.\ (1994) found that for Wolf-Rayet
winds, a typical mass spectrum that follows $m_c^{-1.5}$ provides a
reasonably good description of the emission line variations, where
$m_c \propto \rho_{\rm c}\,l_{\rm c}^{3}$ is the mass of a clump. Taken
together, the above scaling laws favor an asperous wind with clumps of
varying size but constant column density and a mass spectrum
dramatically raising to lower clump masses.

If full-scale turbulence really does prevail, the discrete clumps
could in principle be usable to suppress a substantial fraction of the
ionizing photon flux at $E_\gamma = h\nu > 64$ eV from the
afterglow.  To calculate the properties of the asperous wind, we first
consider the following three constraints. First, the clump column
density should be at least $N({\rm C\,IV})\approx 1.5\times 10^{18}\
{\rm cm}^{-2}$ so that $N({\rm C\,IV})\times \sigma_{ph}({\rm
C}^{3+})=1$ for suppressing a substantial fraction of the ionizing
flux. Because the ionization potentials of He$^+$ and C$^{3+}$ are
very similar (see the inset of Figure 13)\footnote{We note that
photons of $E_{phot} > 350$ eV are energetic enough to remove
electrons from the inner shell (K-shell states) and electrons in the
outer shell (L-shell states) cascade down to fill the vacancy on time
scales of a few fs (Schlachter \etal\ 2004).  This explains the edge
that appears at 350 eV in the inset of Figure 13.  But because the
afterglow radiation declines quickly with frequency ($\beta=-1$),
ionization of C$^{3+}$ due to the removal of a K-shell electron is
limited to $\sim 10$\%.}, we treat He$^{+}$ and C$^{3+}$ as the same
particles in the photo-ionization process. This leads to a minimum gas
clump gas column density of $N_g=N({\rm C\,IV})\times ({\rm C}/{\rm
He})^{-1} \approx 1.5\times 10^{19}\ {\rm cm}^{-2}$ for Wolf-Rayet
winds with ${\rm C}/{\rm He}=0.1$.  Second, the clump size should be
at least comparable to the apparent image size of the afterglow
(e.g. Granot et al.\ 2005),
\begin{equation}
r_\perp = 2 \times 10^{16}\,E_{{\rm iso},52}^{1/4}\,\dot{M}_{-5}^{1/4}\,v_{w,3}^{-1/4}\,[t_{\rm days}/(1+z)]^{3/4}\ {\rm cm}.
\end{equation}
This leads to a minimum clump mean gas density of $\rho_{\rm c,min}
\sim 750$ cm$^{-3}$ and a minimum clump mass of $m_{\rm c,min}\sim 4
\times 10^{-5} M_\odot$. Third, the total mass of the unshocked
Wolf-Rayet wind sets a limit to the total number of available clumps
at the smallest scale (note that given the assumed mass spectrum, the
total mass of the outflow is dominated by the lower mass clumps, here
assumed to be $\sim r_\perp$). In the unshocked wind, the mass within
a radius $r$ is $\dot{M}\,r/v_w$, which combined with the minimum clump
mass estimate gives the maximum number of clumps of size $l_c \sim
r_\perp$ within the $1/r^2$ wind:
\begin{equation}
n_c = 10^{2}\, \dot{M}_{-5}^{7/4}\, v_{w,3}^{-7/4}\, \rho_{\rm c,min}^{-1}\, r_{\rm pc}.
\end{equation}
The volume filling factor is thus given by 
\begin{equation}
f_c = 3 \times 10^{-5}\, \dot{M}_{-5}\, v_{w,3}^{-1}\,\rho_{\rm
c,min}^{-1}\,r_{\rm pc}^{-2}.
\end{equation}

Whether a clumpy wind scenario is a viable solution for the survival
of C$^{3+}$ at $< 30$ pc depends on the number of clumps one can
reasonably find along the line of sight to the afterglow.  Adopting a
total ionizing photon emission of $\phi_\gamma=4\times 10^{60}$ for
GRB\,021004 from Lazzati \etal\ (2006), we estimate a photon flux
accumulated in time of $N_\gamma=3\times 10^{22}$ cm$^{-2}$ at
$r=1$~pc.  For $\rho_{\rm c,min}$ one requires roughly
\begin{equation}
n_\tau =  2\times 10^3\, \rho_{\rm c,min}^{-1}\,r_{\rm pc}^{-2} 
\end{equation}
clumps to be present along the sightline to effectively absorb all the
ionizing photons. It is therefore clear that a clumpy wind model
fails, because there is simply not enough mass in the unshocked wind
medium. A close inspection of the above requirements shows that for a
fixed unshocked wind mass, altering $\rho_{\rm c,min}$, $r_{\rm pc}$
or $l_c$ (here assumed to be $\sim r_\perp$) fails to improve the
situation. For example, in an extreme case scenario a single clump of
$N_g=1.5\times 10^{22}\ {\rm cm}^{-2}$ (total clump mass of 0.05
$M_\odot$ for $l_c=r_\perp$), at $r$=1 pc along the sightline is
sufficient to absorb the large incident ionizing flux, but requires
$\dot{M} \gg 10^{-2} M_\odot {\rm yr}^{-1}$ so that the probability of
having one in front of the afterglow will be at least modest. We
contend that a clumpy wind model is therefore an unlikely solution.

After a careful consideration of various scenario, we conclude that
C$^{3+}$ ions at $r< 30$ pc cannot survive the intense afterglow
radiation field to impose an absorption feature in the afterglow
spectrum.

\section{SUMMARY AND CONCLUSIONS}

We have collected a sample of five GRB afterglow spectra for an
unbiased search of blue-shifted, high-velocity C\,IV absorbers at
$|\Delta\,v|=1000-5000$ \kms\ from the host redshift $z_{\rm GRB}$.
These absorbers (designated C\,IV$_{15}$) are candidate stellar wind
features expected from a Wolf-Rayet progenitor of long-duration GRBs.
Our sample, although small, is statistically unbiased, because it is
assembled based on all available afterglow spectra suitable for our
study without prior knowledge of the line-of-sight properties.  We
identify only one C\,IV$_{15}$ absorber at $\Delta\,v=-1500$ \kms\
from GRB\,050730 and none along the rest of the sightlines to a
3-$\sigma$ limit of EW$= 0.4$ \AA\ over a uniform $\delta\,v=170$
\kms\ velocity resolution element.

Our search yields an estimate of 20\% for the incidence of
C\,IV$_{15}$ absorbers from the GRB host galaxies with a 68\%
confidence interval of 0--40\%.  This is consistent with what is
expected for classical damped \lya\ absorbers toward quasar
sightlines.  The result suggests that the majority of these
C\,IV$_{15}$ absorbers originate in a foreground galaxy along the
sightline.

We have also assembled an early sample based on GRB sightlines from
earlier work, including GRBs 000926 (Castro \etal\ 2003), 021004
(Mirabal \etal\ 2003; Fiore \etal\ 2005), and 030323 (Vreeswijk \etal\
2004).  The sightline toward GRB\,021004 is the only source that
exhibits two C\,IV$_{15}$ absorbers at $\Delta\,v=-2675$ \kms\
(Component 1) and $\Delta\,v=-2900$ \kms\ (Component 2), respectively,
as reported by previous authors.  The C\,IV$_{15}$ absorbers found in
the early sample offer additional insights for understanding their
origin.

The presence of a saturated C\,II $\lambda\,1334$ absorption and
absence of C\,II* $\lambda\,1335$ transition together allow us to rule
out the possibility of Component 1 originating in the GRB progenitor
environment.  Our analysis shows that had the gas originated in the
vicinity of the afterglow (i.e.\ at $r < 1$\,kpc), it would have been
highly excited by the UV photons of the afterglow.  Instead, Component
1 most likely originates in a foreground galaxy along the line of
sight, which is supported by both the statistical expectation of the
the incidence of C\,IV absorbers and the presence of a faint galaxy at
0.3\arcsec\ away from the GRB host galaxy.

The lack of wind signatures for 80\% of our statistical sample is
understood as due to gas being photo-ionzed by the intense UV
radiation field of the afterglows.  Our estimates show that ionizing
photons from the afterglow can ionize C$^{3+}$ to beyond 30 pc radius
from the progenitor stars.  We explore different scenarios that would
allow C$^{3+}$ to survive at close distances from the afterglow.  We
find based on a chemical composition C/He=0.1, typical of local
Wolf-Rayet winds, a large density contrast in the wind can in
principle provide necessary shielding for the C$^{3+}$ ions, but but
requires $\dot{M} \gg 10^{-2} M_\odot {\rm yr}^{-1}$ so that the
probability of having one in front of the afterglow will be at least
modest. We contend that a clumpy wind model is unlikely to be a viable
scenario.

\acknowledgments
 
 We thank N.\ Mirabal and P.\ Vreeswijk for sharing their spectra of
GRB\,021004 and GRB\,030323.  H.-W.C. acknowledges helpful discussions
with A.~J. van Marle, P. Crowther, J. Eldridge, C. Chin, and
A. K{\"o}nigl.  We thank Davide Lazzati and an anonymous referee for
critical comments that helped to improve the paper.  H.-W.C., JXP, and
JSB acknowledge partial support from NASA grant NNG05GF55G.
H.-W.C. was partially supported by NASA grant NNG06GC36G and NSF grant
AST-0607510. E.R-R acknowledges support from the John Bahcall
Fellowship.

%\newpage

\newpage

\begin{figure*}[bht]
\begin{center}
\includegraphics[scale=0.6,angle=0]{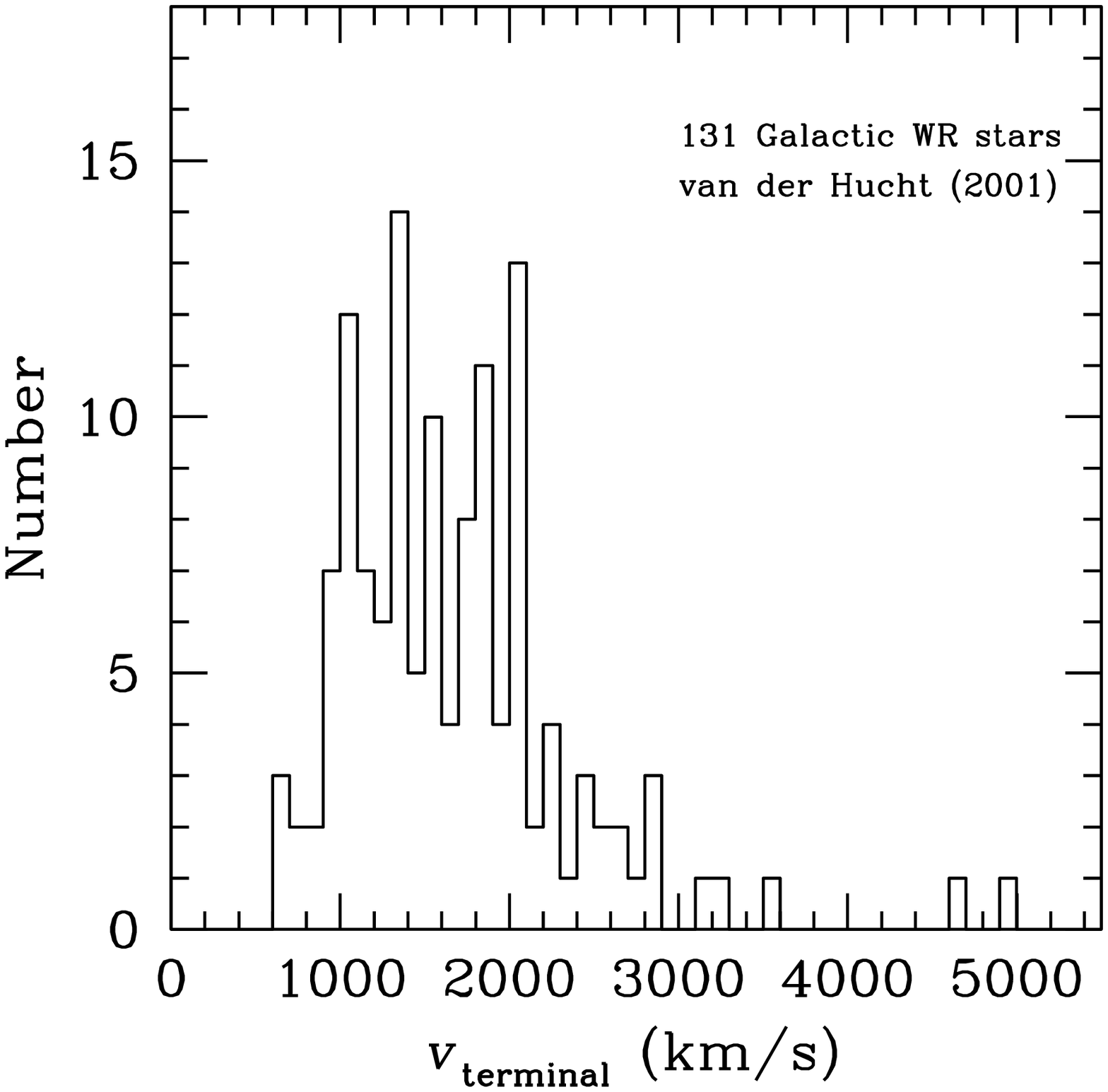}
\end{center}
\caption[]{Number counts of Galactic Wolf-Rayet stars versus
terminal wind velocity as determined from P-Cygni profiles of
absorption lines like C\,IV $\lambda\,1550$ (van der Hucht 2001).}
\end{figure*}

\begin{figure*}
\begin{center}
\includegraphics[scale=0.7,angle=270]{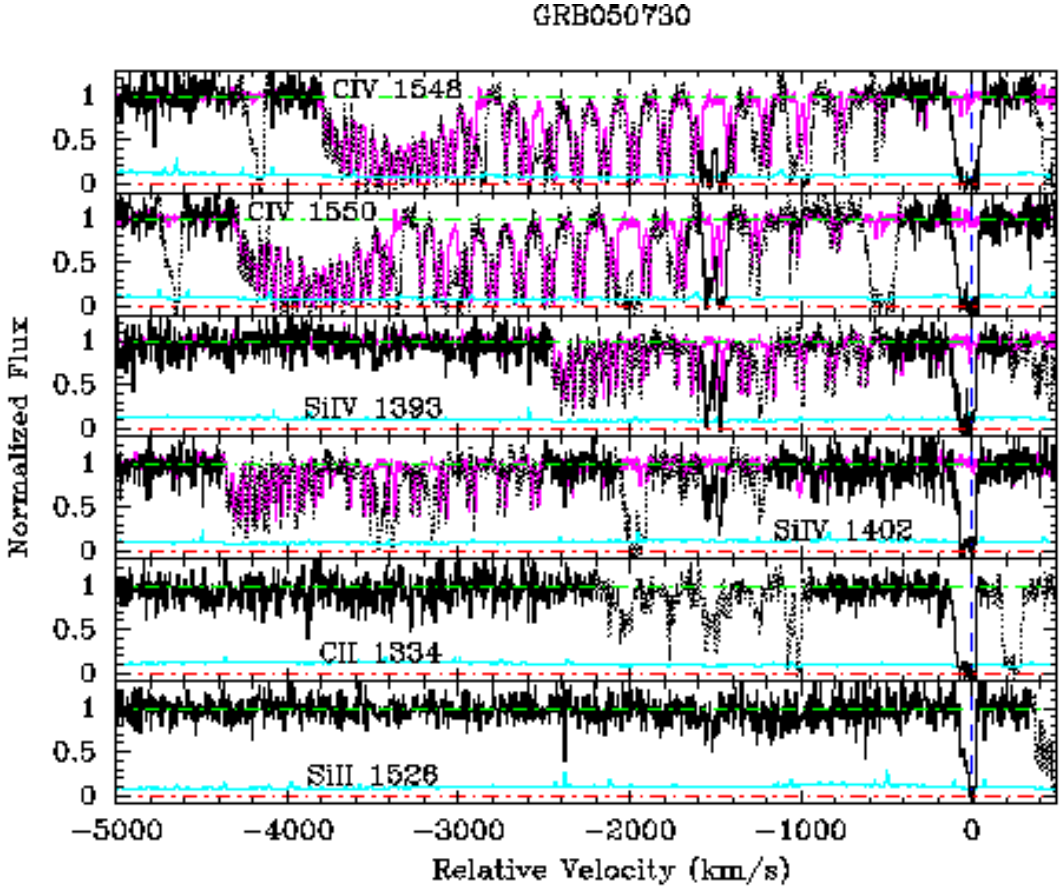}
\end{center}
\caption[]{Absorption profiles of C\,IV $\lambda\lambda\,1548, 1550$,
  Si\,IV $\lambda\lambda\,1393, 1402$, C\,II $\lambda\,1334$, and
  Si\,II $\lambda\,1526$ observed in the host of GRB\,050730.  Zero
  relative velocity corresponds to $z=3.96855$.  Over the velocity
  interval from $\Delta\,v=-5000$ \kms\ through $\Delta\,v=-1000$
  \kms, we detect one blue-shifted C\,IV and Si\,IV absorber at
  $\Delta\,v\approx -1500$ \kms.  The CIV features are contaminated by
  the atmosphere A-band absorption forest.  Although we are unable to
  obtain an accurate measurement of the column density of the $C^{3+}$
  ions, we place a lower limit to the rest-frame absorption equivalent
  width of EW$(\lambda\,1550)>0.36$ \AA.  The detection of the C\,IV
  absorption doublet is robust from the comparison with a quasar
  spectrum (the magenta curve) obtained using the same instrument at
  the same observation site (PKS2000; Prochter et al.\ 2007, in
  preparation).  For Si\,IV, we measure $\log\,N({\rm Si\,IV})=13.9\pm
  0.1$.  The contaminating absorption features are dotted out.  The
  dashed-dotted lines indicate the normalized continuum and zero level
  for guide.  The 1-$\sigma$ error array is shown in thin, cyan line.
  Zero relative velocity corresponds to $z=2.32897$.}
\end{figure*}

\begin{figure*}
\begin{center}
\includegraphics[scale=0.6,angle=270]{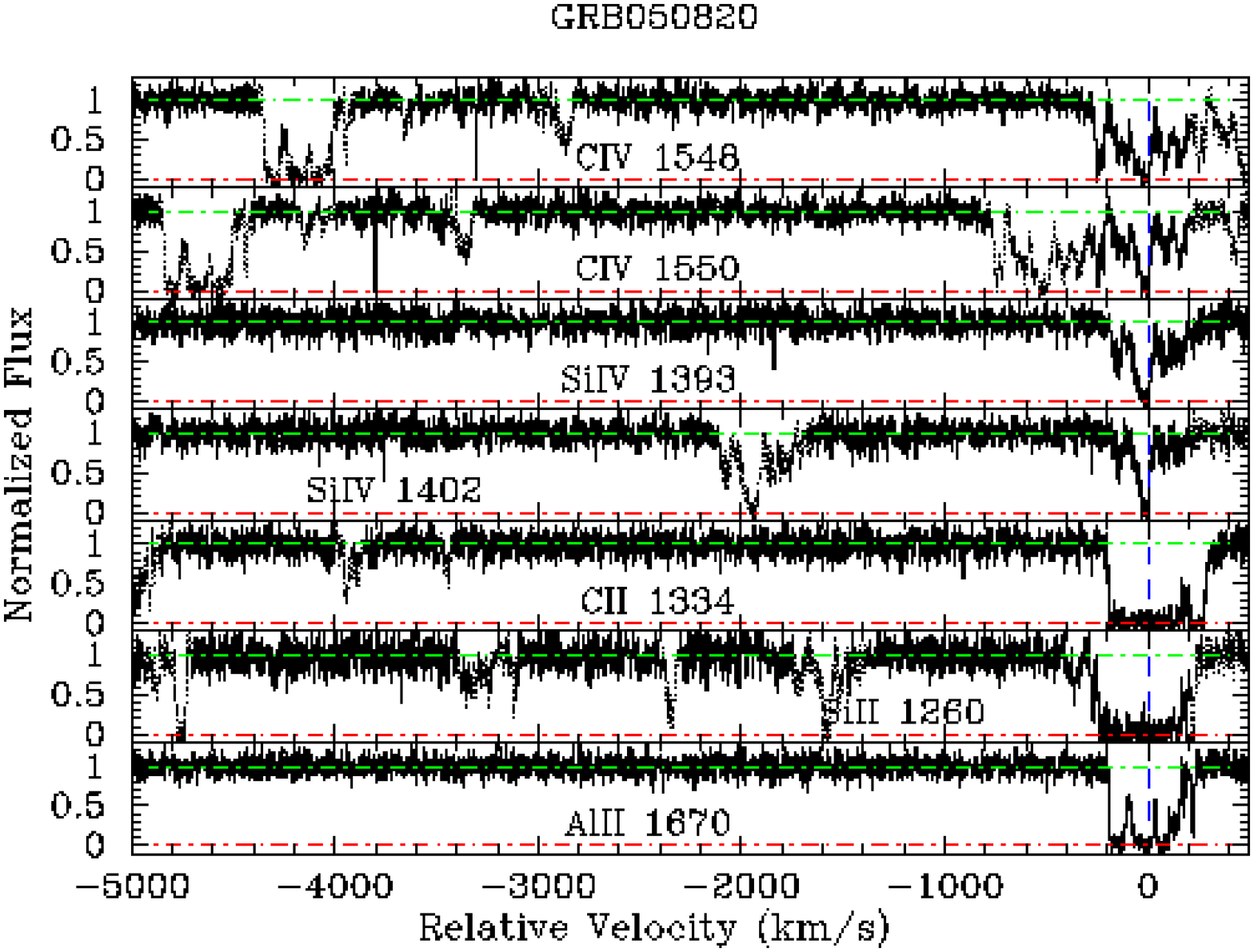}
\end{center}
\caption[]{Absorption profiles of C\,IV $\lambda\lambda\,1548, 1550$,
  Si\,IV $\lambda\lambda\,1393, 1402$, C\,II $\lambda\,1334$, Si\,II
  $\lambda\,1260$, and Al\,II $\lambda\,1670$ observed in the host of
  GRB\,050820.  Zero relative velocity corresponds to $z=2.6147$.
  Over the velocity interval from $\Delta\,v=-5000$ \kms\ through
  $\Delta\,v=-1000$ \kms, we do not detect additional CIV features
  with a 3-$\sigma$ upper limit in the rest-frame absorption equivalent
  width of EW$=0.2$ \AA\ over the spectral resolution element of
  $\delta\,v=7.5$ \kms.}
\end{figure*}

\begin{figure*}
\begin{center}
\includegraphics[scale=0.7,angle=270]{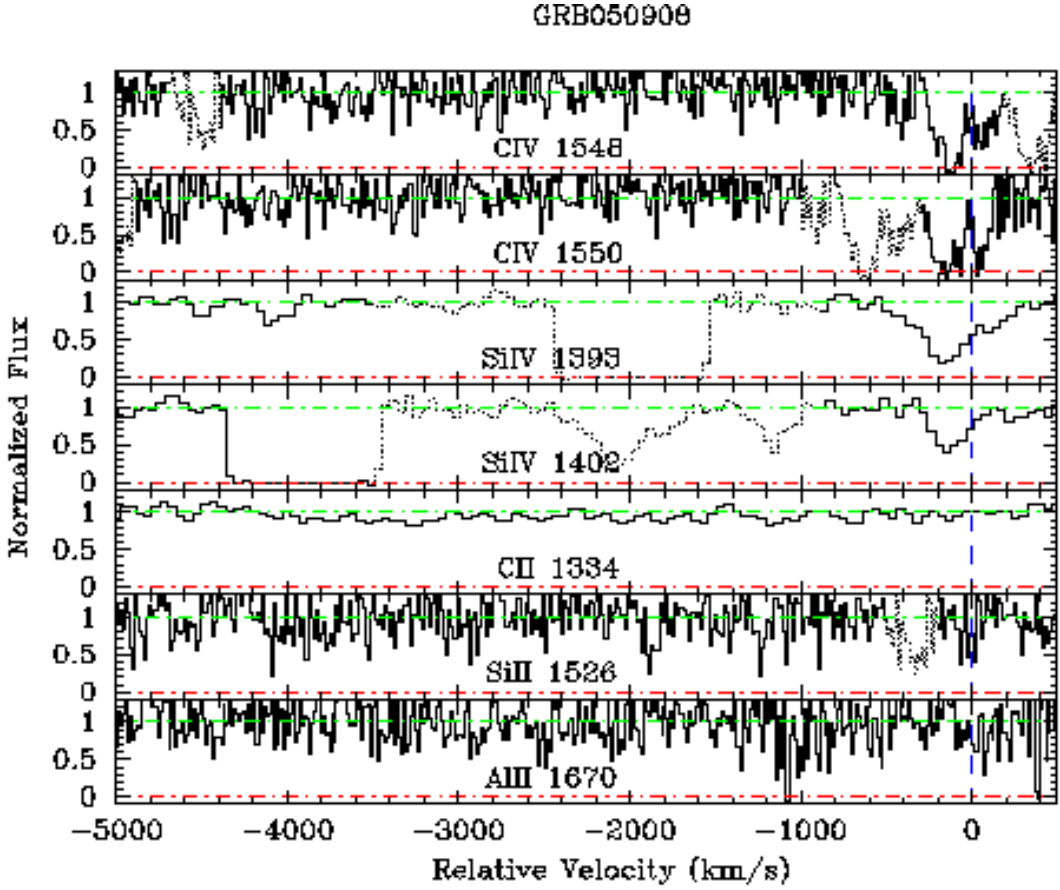}
\end{center}
\caption[]{Absorption profiles of C\,IV $\lambda\lambda\,1548, 1550$,
  Si\,IV $\lambda\lambda\,1393, 1402$, C\,II $\lambda\,1334$, Si\,II
  $\lambda\,1526$, and Al\,II $\lambda\,1670$ observed in the host of
  GRB\,050908.  Zero relative velocity corresponds to $z=3.344$.
  High-resolution data were taken using DEIMOS on the Keck II
  telescope and low-resolution data were taken using GMOS on the
  Gemini north telescope, which had non-contiguous spectral coverage
  due to chip gaps in the camera.  One of the chip gaps is apparent in
  the Si\,IV doublet transitions.  Over the velocity interval from
  $\Delta\,v=-5000$ \kms\ through $\Delta\,v=-1000$ \kms, we do not
  detect additional CIV features with a 3-$\sigma$ upper limit in the
  rest-frame absorption equivalent width of EW$(\lambda\,1548)=0.28$
  \AA\ over the spectral resolution element of $\delta\,v=40$ \kms.}
\end{figure*}

\begin{figure*}
\begin{center}
\includegraphics[scale=0.7,angle=270]{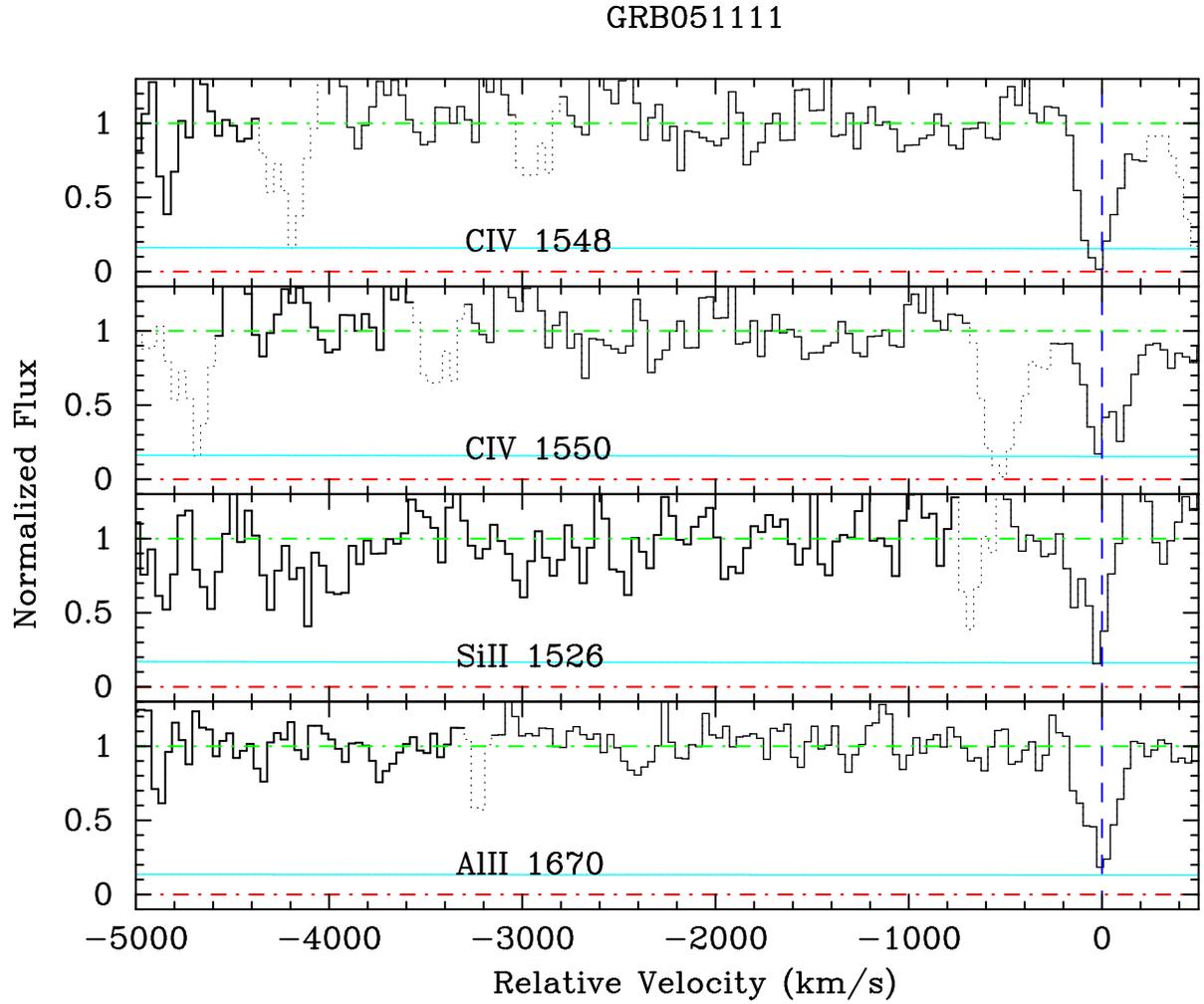}
\end{center}
\caption[]{Absorption profiles of C\,IV $\lambda\lambda\,1548, 1550$,
  Si\,II $\lambda\,1526$, and Al\,II $\lambda\,1670$ observed in the
  host of GRB\,051111 with GMOS on Gemini-N.  
  Zero relative velocity corresponds to
  $z=1.54948$.  Over the velocity interval from $\Delta\,v=-5000$
  \kms\ through $\Delta\,v=-1000$ \kms, we do not detect additional
  CIV features with a 3-$\sigma$ upper limit in the rest-frame
  absorption equivalent width of EW$(\lambda\,1548)=0.37$ \AA\ over
  the spectral resolution element of $\delta\,v=170$ \kms.}
\end{figure*}

\begin{figure*}
\begin{center}
\includegraphics[scale=0.7,angle=270]{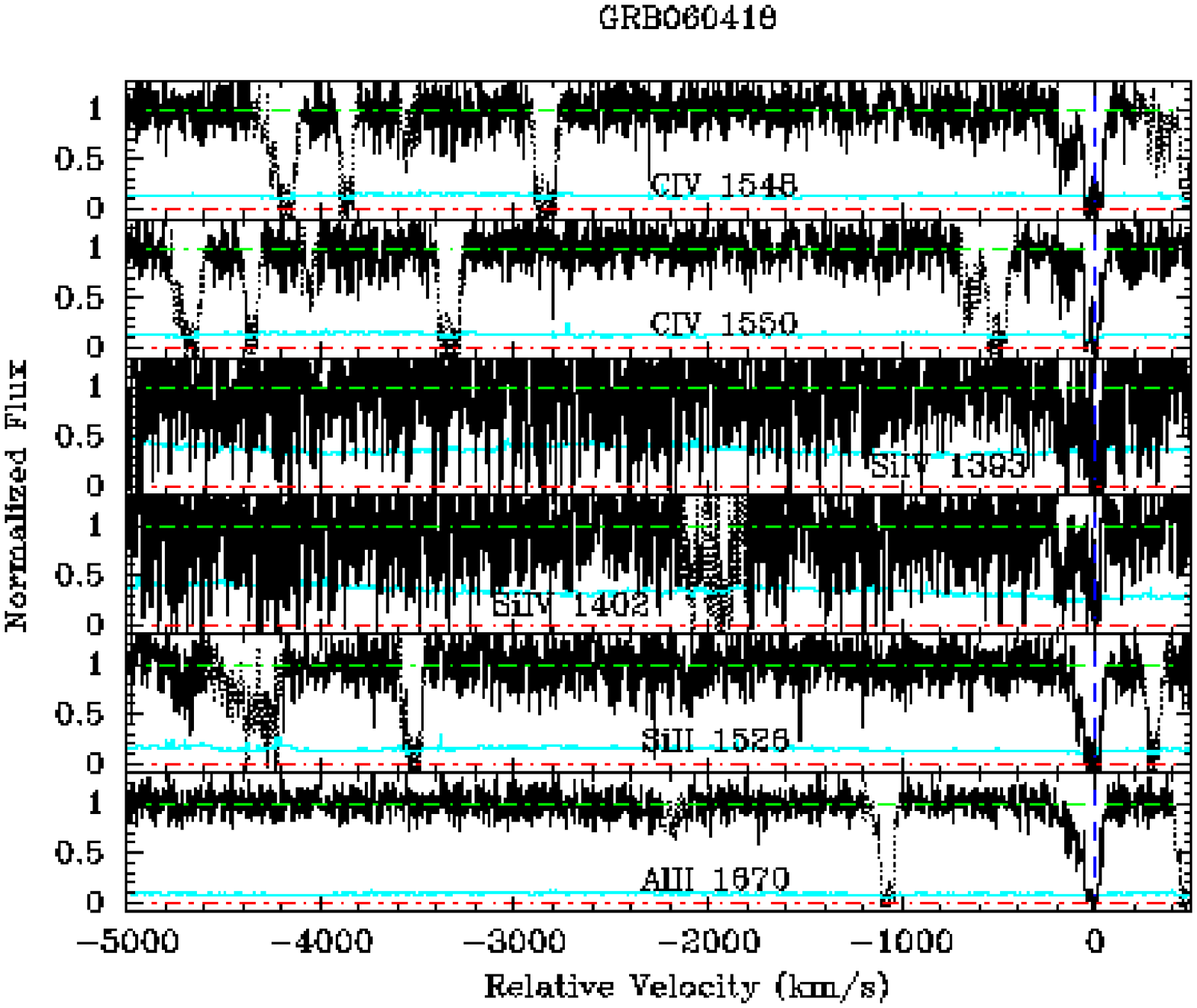}
\end{center}
\caption[]{Absorption profiles of C\,IV $\lambda\lambda\,1548, 1550$,
  Si\,IV $\lambda\lambda\,1393, 1402$, C\,II $\lambda\,1334$, Si\,II
  $\lambda\,1526$, and Al\,II $\lambda\,1670$ observed in the host of
  GRB\,060418.  Zero relative velocity corresponds to $z=1.4901$.
  Over the velocity interval from $\Delta\,v=-5000$ \kms\ through
  $\Delta\,v=-1000$ \kms, we do not detect additional CIV features
  with a 3-$\sigma$ upper limit in the rest-frame absorption equivalent
  width of EW$(\lambda\,1548=0.36$ \AA\ over the spectral resolution element of
  $\delta\,v=12$ \kms.}
\end{figure*}

\begin{figure*}
\begin{center}
\includegraphics[scale=0.7,angle=270]{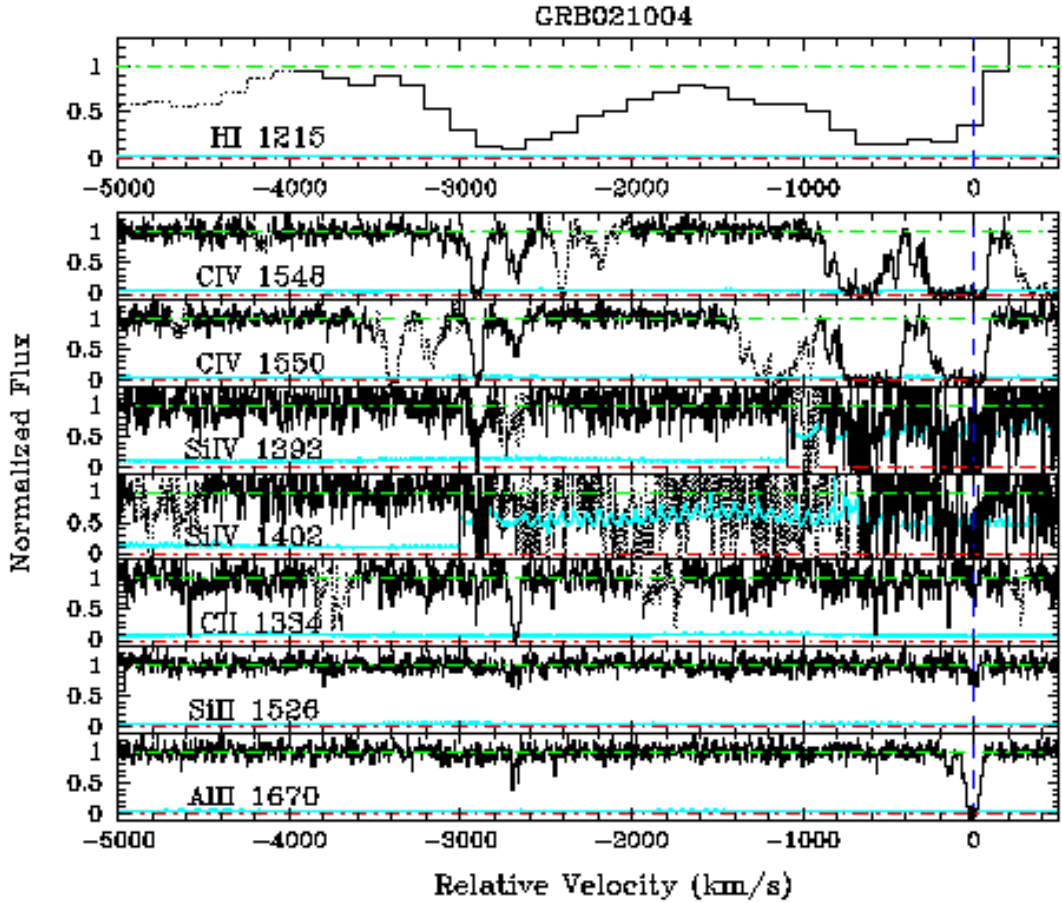}
\end{center}
\caption[]{Absorption profiles of H\,I \lya\ $\lambda\,1215$, C\,IV
  $\lambda\lambda\,1548, 1550$, Si\,IV $\lambda\lambda\,1393, 1402$,
  C\,II $\lambda\,1334$, Si\,II $\lambda\,1526$, and Al\,II
  $\lambda\,1670$ observed in the host of GRB\,021004.  The same as
  Figure 2, where contaminating absorption features are dotted out.
  The dashed-dotted lines indicate the normalized continuum and zero
  level for guide.  The 1-$\sigma$ error array is shown in thin, cyan
  line.  Zero relative velocity corresponds to $z=2.32897$.  Over the
  velocity interval from $\Delta\,v=-5000$ \kms\ through
  $\Delta\,v=-1000$\kms, we identify two absorption components at
  $\Delta\,v=-2675$\kms\ and $\Delta\,v=-2900$\kms, respectively.}
\end{figure*}

\begin{figure*}
\begin{center}
\includegraphics[scale=0.7,angle=0]{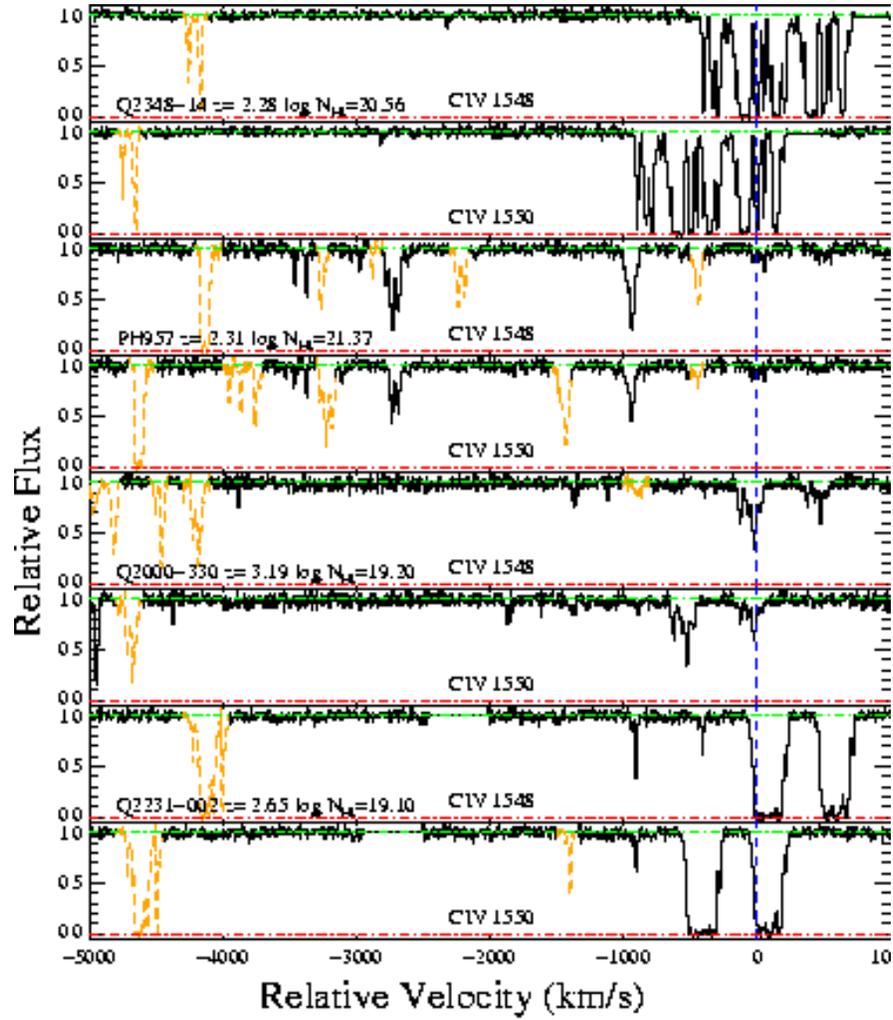}
\end{center}
\caption[]{A sample of four strong intervening \lya\ absorbers, two
DLAs and two super Lyman limit systems, at $2.2 < z < 3.2$ toward
quasar lines of sight.  Contaminating features have been dotted out.
These are part of the 53 quasar sightlines from the Keck/UCSD High
Resolution database (Prochaska \etal\ 2007).  We adopt
this sample of 53 quasars as a control sample for estimating the
statistical significance of possible ``over-abundance'' of C\,IV$_{15}$
absorbers.}
\end{figure*}

\begin{figure*}
\begin{center}
\includegraphics[scale=0.7,angle=0]{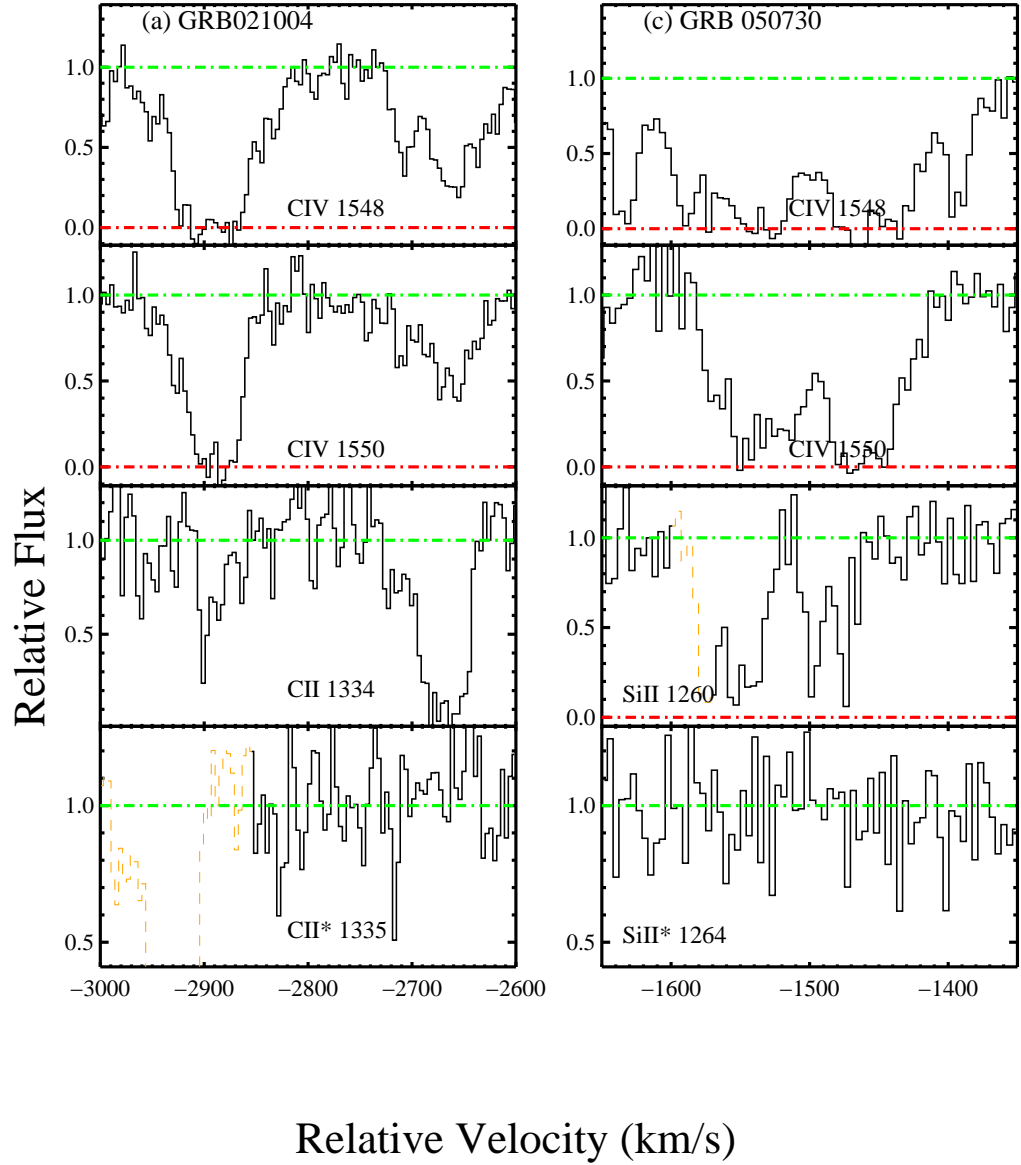}
\end{center}
\caption[]{Velocity profiles for the C\,IV$_{15}$ absorbers from
GRB~021004 (left panels) and from GRB~050730 (right panels).  We also
present the low-ion transitions of C\,II $\lambda$1334 and
Si\,II$\lambda$1260 for GRB\,021004 and GRB\,050730, respectively,
together with their associated fine-structure transitions.  No excited
transitions are detected in either of the sources.}
\end{figure*}

\begin{figure*}
\begin{center}
\includegraphics[scale=0.5,angle=270]{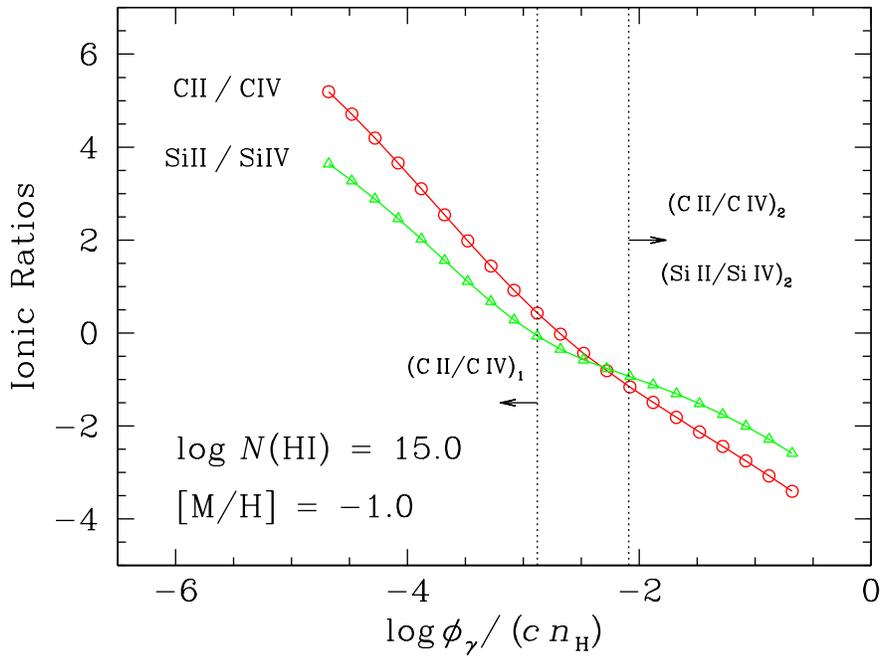}
\end{center}
\caption[]{Expected ionic abundance ratios versus the ionization
parameter $U\equiv \phi_{\gamma}/c\,n_{\rm H}$.  The expectations are
calculated using the Cloudy software (Ferland \etal\ 1998; version
06.02) for clouds of plane parallel geometry and under
photo-ionization equilibrium.  We have assumed a metallicity of 0.1
solar and $\log\,N(\hI)=15$ and adopted the background radiation field
of Haardt \& Madau (2006 in preparation) that includes ionizing
photons from both quasars and starburst galaxies.  Constraints on the
ionization parameter based on the observed relative abundances between
different ionization stages are shown in vertical lines for the two
components toward GRB\,021004.}
\label{fig:cloudy}
\end{figure*}

\begin{figure*}[h]
\begin{center}
\includegraphics[scale=0.75,angle=0]{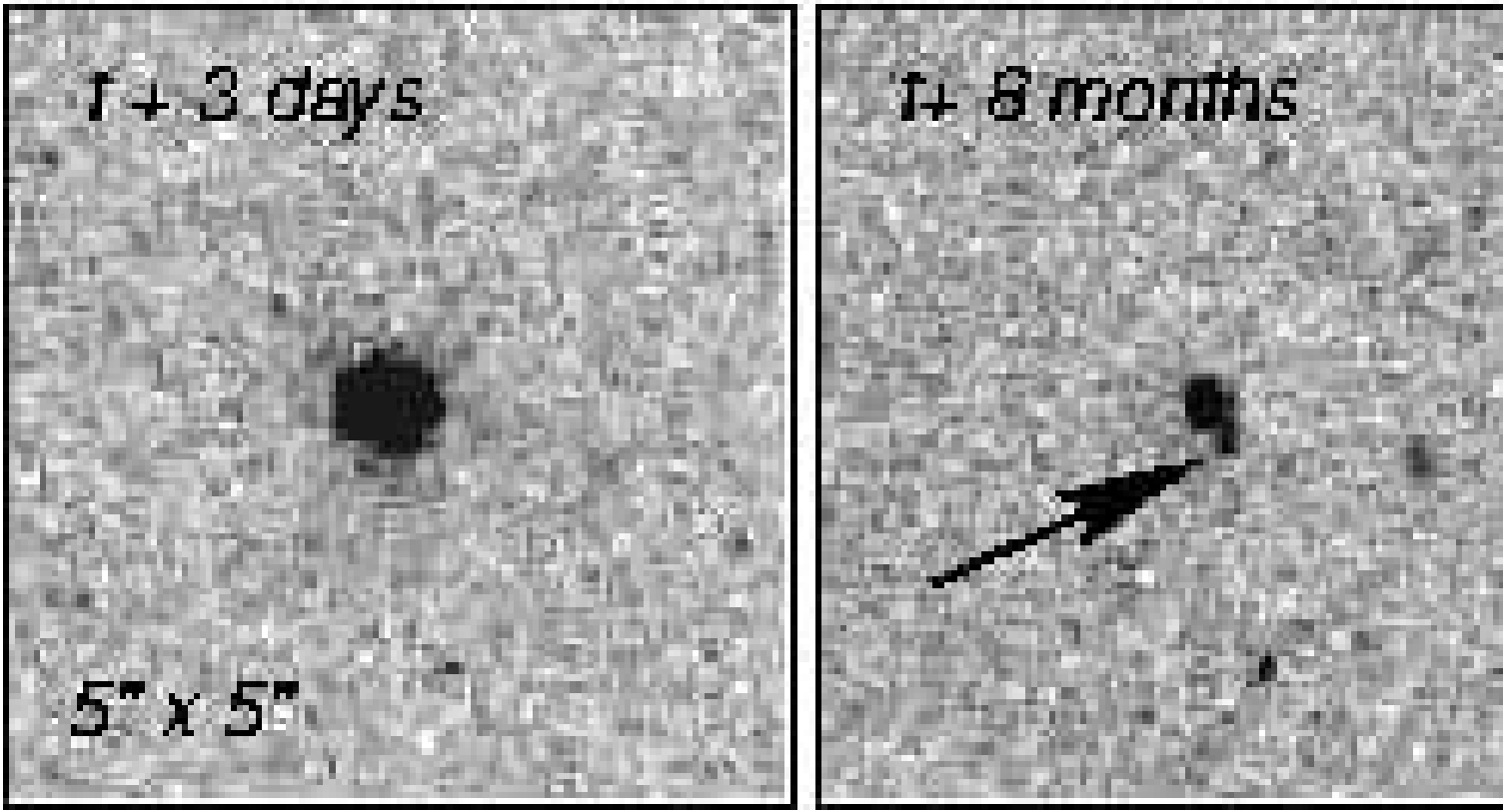}
\end{center}
\caption[]{Images of the field surrounding GRB\,021004, obtained using
HST/ACS with the F606W filter three days (left panel) and eight months
(right panel) after the trigger.  The second epoch image (right
panel), obtained when the optical afterglow had faded, clearly shows a
faint companion (pointed by the arrow) next to the host galaxies at
the center of the frame. }
\label{fig:img021004}
\end{figure*}

\begin{figure*}
\begin{center}
\includegraphics[scale=1.0,angle=0]{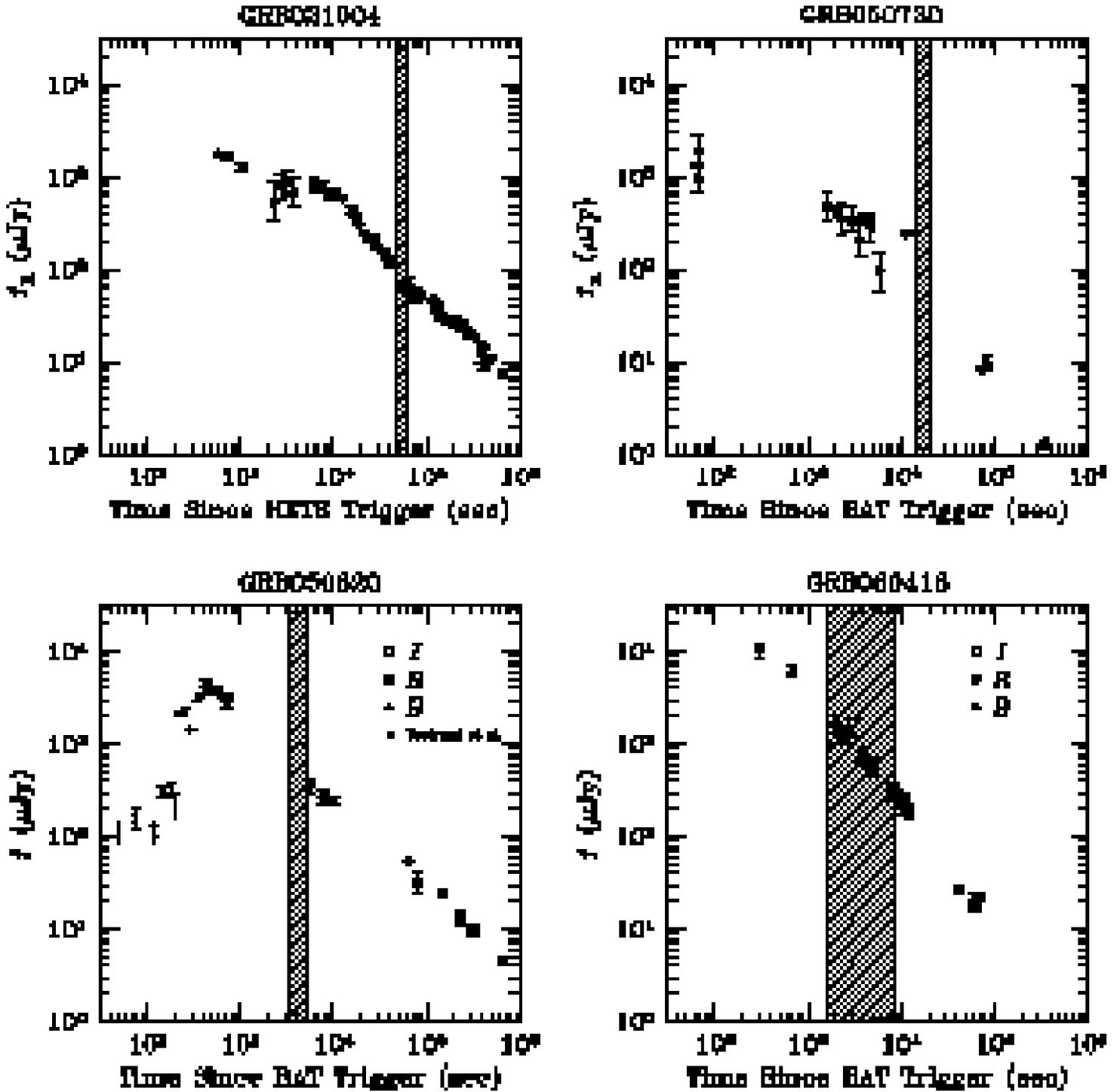}
\end{center}
\caption[]{Optical light curves of GRBs 021004, 050730,
050820, and 060418, for which photometric measurements from $<$ 10
minutes after the initial burst are available.  The vertical shaded
regions indicate the period when the spectroscopic data were taken.
The light curves of GRBs 021004 and 050730 are based on $R$-band
observations alone, while the light curves of GRBs 050820 and 060418
include observations in the $B$, $R$, and $I$ bands.  We have
corrected for the Galactic extinction, when combining photometry from
different bandpasses with $E(B-V)=0.04$ toward GRB\,050820 and
$E(B-V)=0.22$ toward GRB\,060418.  Observations of GRB021004 were
taken from Holland \etal\ (2003) and Uemura \etal\ (2003).
Observations of GRB\,050730 were taken from Klotz \etal\ (2005) and
Damerdji \etal\ (2005).  Observations of GRB\,050820 were taken from
Vestrand \etal\ (2006).  Observations of GRB\,060418 were taken from
Schady \& Falcone (2006) and Cobb (2006). }
\label{fig:lcurve}
\end{figure*}

\begin{figure*}
\begin{center}
\includegraphics[scale=0.9,angle=0]{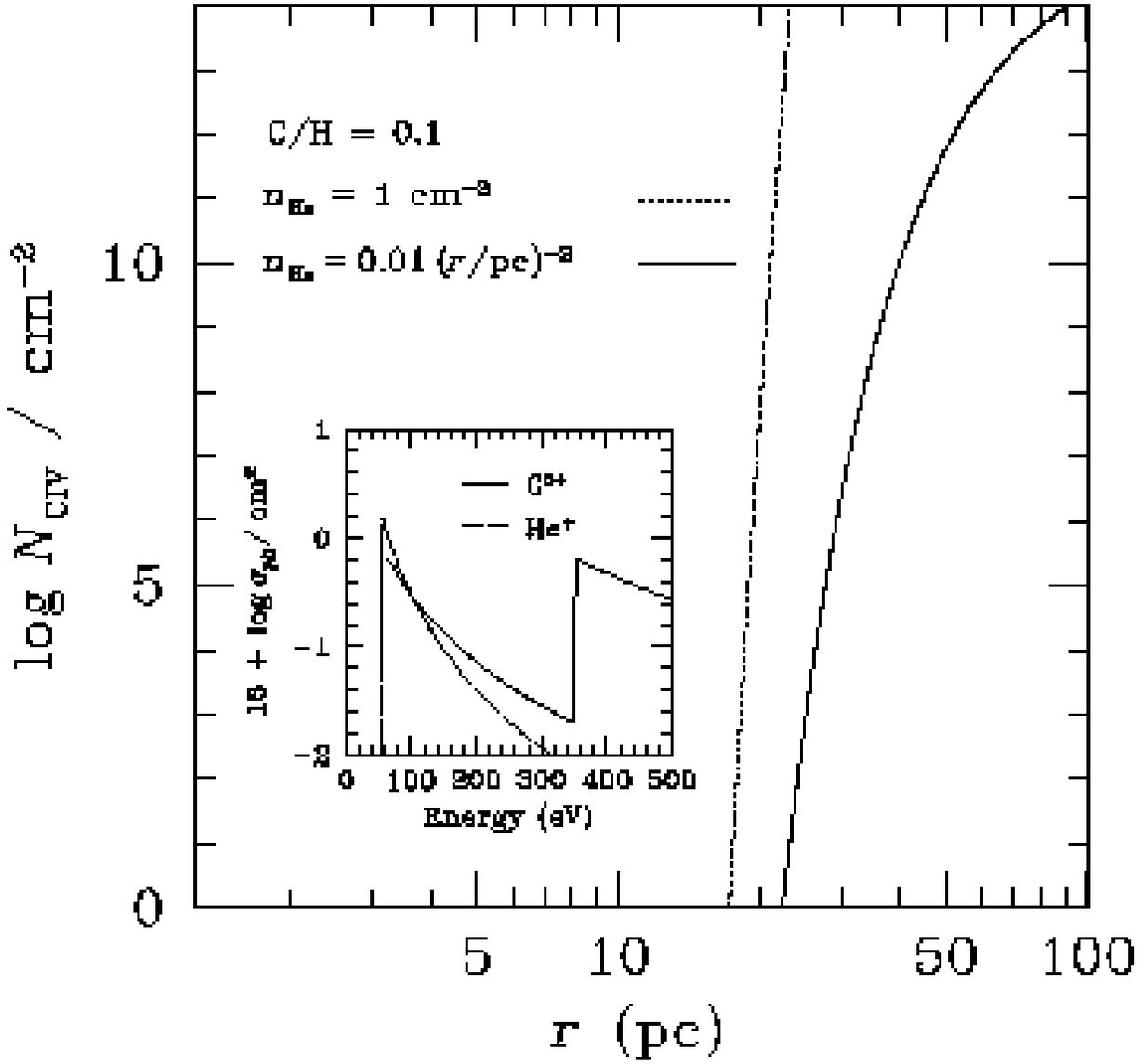}
\end{center}
\caption[]{Estimated $N({\rm C\,IV})$ versus distance to the
afterglow in the presence of He$^+$ for two different density
profiles.  The inset shows the photo-ionization cross sections of
He$^+$ and C$^{3+}$ versus photon energy.  The curves are calculated
according to Verner \& Yakovlev (1995) and Verner \etal\ (1996).}
\end{figure*}

\begin{deluxetable}{p{0.8in}ccccrc}
\tabletypesize{\scriptsize}
%\rotate
\tablewidth{0pt}
\tablecaption{Summary of Absorption-line Properties}
\tablehead{\multicolumn{1}{c}{Field} & \colhead{$E_{\rm iso} (\times 10^{53}$ ergs)\tablenotemark{a}} & \colhead{$z_{\rm GRB}$} & \colhead{$\log\,N(\hI)$} & \colhead{$[\frac{\rm M}{\rm H}]_{\rm ISM}$} & \colhead{$-\Delta\,v_{\rm C\,IV}^{\rm max}$} & \multicolumn{1}{c}{$t_{\rm obs}^{\rm spec}$ (hr)}} 
\startdata
\multicolumn{7}{c}{The Statistical Sample} \nl
\tableline
%GRB\,050505 \dotfill & $2.3\pm 0.5$ & 4.274 & $22.05\pm 0.10$ & $>-0.1$ & $1000$ &  ($6.5-7$ \nl
GRB\,050730 \dotfill & $1.2\pm 0.2$ & 3.968 & $22.15\pm 0.05$ & $-2.0\pm 0.1$ & $1500$\tablenotemark{b} & ($4.0-5.7$) \nl
GRB\,050820A \dotfill & $2.0\pm 0.4$ & 2.615 & $21.00\pm 0.10$ & $-0.6\pm 0.1$ & 250 & ($0.9-1.4$) \nl
GRB\,050908 \dotfill & $0.11\pm 0.01$ & 3.344 & $19.2\pm 0.2$ & $\sim\,-0.9$ & 200 & ($3.7-4.3$)/($7.7-8.4$)\tablenotemark{c} \nl
GRB\,051111 \dotfill & $0.62\pm 0.05$ & 1.549 & $>20.3$ & $< 0.9$ & 200 & ($2.6-3.3$) \nl
GRB\,060418 \dotfill & $0.67\pm 0.05$ & 1.490 & $>20.3$ & $< 0.0$ & 250  & ($0.5-2.3$) \nl
\tableline
\multicolumn{7}{c}{The Early Sample} \nl
\tableline
GRB\,000926 \dotfill & $2.4\pm 0.8$ & 2.038 & $21.3\pm 0.2$ & $-0.17\pm 0.05$ & 250 & ($29.1-30.5$) \nl
GRB\,021004 \dotfill & $0.5\pm 0.1$ & 2.328 & $19.6\pm 0.3$ & $-1.3\pm 0.5$ & $2900$ & ($13.6-17.1$)  \nl
GRB\,030323 \dotfill & $0.3\pm 0.1$ & 3.372 & $21.90\pm 0.07$ & $-1.0\pm 0.2$ & $200$ & ($57.2-61.2$)  \nl
\enddata
\tablenotetext{a}{GRBs 050730, 050820A, 050908, 051111 \& 060418: Butler \etal\ 2006, in preparation.  GRB\,000926: Harrison \etal\ 2001; GRBs 021004 \& 030323: Sakamoto \etal\ 2005. }
\tablenotetext{b}{Contaminated by the atmosphere A-band absorption.}
\tablenotetext{c}{Spectra obtained using GMOS on the Gemini North telescope and DEIMOS on the Keck telescope, respectively.  See descriptions in \S\ 2.6.}
\end{deluxetable}

%\clearpage

%\begin{deluxetable}{p{1.0in}cccccc}
%\tabletypesize{\normalsize}
%\rotate
%\tablewidth{0pt}
%\tablecaption{Ionic Column Densities of High-velocity Gas}
%\tablehead{ \colhead{} & \colhead{} & \multicolumn{2}{c}{GRB021004} & \colhead{} & \multicolumn{2}{c}{GRB050730} \\
%\cline{3-4} \cline{6-7} \\
%\multicolumn{1}{c}{Transition} &\colhead{} & \colhead{$\Delta\,v=-2900$ \kms} & \colhead{$\Delta\,v=-2675$ \kms} & \colhead{} & \colhead{$\Delta\,v=-1543$ \kms} & \colhead{$\Delta\,v=-1469$ \kms}}
%\startdata
%C\,IV 1550 \dotfill & & $>14.72$ & $14.07\pm 0.02$ & & contaminated & contaminated \nl
%C\,II 1334 \dotfill & & $13.62\pm 0.04$ & $>14.46$ & & contaminated & contaminated \nl
%Si\,IV 1402 \dotfill & & $13.8\pm 0.6$ & contaminated & & $13.59\pm 0.04$ & $13.64\pm 0.04$ \nl
%Si\,II 1526 \dotfill & & $< 12.92$ & $13.42\pm 0.04$ & & $<13.56$ & $<13.10$ \nl
%\enddata
%\end{deluxetable}

\begin{deluxetable}{p{1.0in}cc}
\tabletypesize{\normalsize}
%\rotate
\tablewidth{0pt} 
\tablecaption{Ionic Column Densities of ``High-velocity'' Gas Toward GRB\,021004\tablenotemark{a,b}}
\tablehead{\colhead{} & \colhead{component 1} & \colhead{component 2} \\ 
\multicolumn{1}{c}{Transition} & \colhead{$\Delta\,v=-2675$ \kms} & \colhead{$\Delta\,v=-2900$ \kms}} 
\startdata 
C\,IV 1550 \dotfill & $14.07\pm 0.02$ & $>14.72$ \nl 
C\,II 1334 \dotfill & $>14.46$ & $13.62\pm 0.04$ \nl 
Si\,IV 1402 \dotfill & contaminated & $13.8\pm 0.3$ \nl 
Si\,II 1526 \dotfill & $13.42\pm 0.04$ & $< 12.92$ \nl
\enddata 
\tablenotetext{a}{We note that Fiore et al. (2005) presented
independent measurements for these transitions.  However, they did not
account for the saturation of strong lines and presented lower limits
as in measurements with errors.}
\tablenotetext{a}{Errors presented here account for only statistical errors. 
Uncertainties due to continuum fitting contribute to $\sim 0.05$ dex.}
\end{deluxetable}

\end{document}